\documentclass[showpacs,superscriptaddress, floatfix, aps, prb,notitlepage, reprint]{revtex4-2}
\usepackage{bm}
\usepackage{amsmath}
\usepackage{amsfonts}
\usepackage{dsfont}
\usepackage{graphics}
\usepackage{braket}
\usepackage[colorlinks, allcolors=blue]{hyperref}
\usepackage[caption=false]{subfig}
\usepackage{overpic}
\usepackage[capitalise]{cleveref}
\usepackage{docmute}

\usepackage[normalem]{ulem}

\usepackage{xcolor}

\newif\ifshowchanges
\showchangestrue       

\ifshowchanges
    \newenvironment{revision}{\begingroup\color{blue}}{\endgroup}
\else
    \newenvironment{revision}{}{}
\fi

\begin{document}
\title{Numerical and theoretical studies of the Anderson transition in the chiral symmetry classes with weak topology}

\author{Shiyin Kuang}
\email{kuangshy@stu.pku.edu.cn}
\affiliation{International Center for Quantum Materials, Peking University, Beijing 100871, China}
\author{Tong Wang}
\affiliation{Max Planck Institute for the Science of Light, 91058 Erlangen, Germany}
\affiliation{International Center for Quantum Materials, Peking University, Beijing 100871, China}
\author{Zhenyu Xiao}
\affiliation{Princeton Quantum Initiative, Princeton University, Princeton, NJ 08544, USA}
\affiliation{International Center for Quantum Materials, Peking University, Beijing 100871, China}
\author{Pengwei Zhao}
\email{pwzhao@stu.pku.edu.cn}
\affiliation{International Center for Quantum Materials, Peking University, Beijing 100871, China}
\author{Ryuichi Shindou}
\email{rshindou@pku.edu.cn}
\affiliation{International Center for Quantum Materials, Peking University, Beijing 100871, China}
\date{\today}

\begin{abstract}
     Combining lattice-model simulations with a renormalization group (RG) analysis of effective theories, we investigate the nature of the Anderson transition in chiral symmetry classes with one-dimensional (1D) weak topology. In the simulation study, we extend previous transfer matrix analyses to the chiral symplectic class, and study numerical Lyapunov exponents via a finite-size scaling (FSS) analysis that assumes spatially isotropic scaling. The analysis shows that, as in the other two chiral symmetry classes, the weak topology induces an intermediate quasi-localized (QL) phase between metal and Anderson insulator phases. In this QL phase, the localization length of wave functions diverges exclusively along the direction of the 1D weak topology. A comparison of the universal critical exponents across the three chiral symmetry classes indicates that the quantum criticality of the metal-to-QL transition varies, depending on the nature of the time-reversal symmetry. Motivated by this numerical observation, we revisit and extend our previous two-dimensional (2D) RG analysis to all three chiral classes, now newly incorporating a one-loop renormalization of the weak topological term in the analysis. The revised analysis reveals that a quasi-localized strong-coupling fixed point previously reported in the chiral unitary class~\cite{zhao2024} is unstable under this new inclusion; instead, the strong-coupling phase is entirely governed by a stable fixed point with conventional localized character. Nevertheless, in the chiral unitary and chiral symplectic classes, the RG analysis still yields the hallmark of the 1D weak topology through the spatially anisotropic scaling of the Anderson transition criticality. These theoretical findings suggest that the quasi-localized phase observed numerically in 2D models may be an artifact of the spatially isotropic scaling assumption in the FSS analysis. A conclusive numerical identification of this phase therefore requires a finite-size scaling approach that accommodates generic (anisotropic) spatial scaling. 

\end{abstract}

\maketitle
\section{Introduction}

    The exploration of novel quantum phases and phase transitions constitutes a central theme in modern physics. According to the celebrated scaling theory of phase transition~\cite{kadanoff1966,wilson1971,wilson1972,abrahams1979,cardy1996}, continuous phase transitions in vastly different physical systems can belong to the same universality class, sharing the same critical properties such as the critical exponents and the scaling functions. Since Altland and Zirnbauer (AZ) established the tenfold symmetry classifications~\cite{altland1997}, it has been widely acknowledged that the universality classes of Anderson transitions are determined exclusively by the symmetry and dimensionality of the disordered Hamiltonians~\cite{wegner1979,efetov1980,hikami1980,hikami1981,efetov1983,mackinnon1981,mackinnon1983,altland1997,slevin1999,evers2008,slevin2014}. Among the ten symmetry classes in the AZ classification, the chiral symmetry classes~\cite{gade1991,gade1993,balents1997,hatsugai1997,altland1997,motrunich2002,horovitz2002,mudry2003,bocquet2003,garcia-garcia2006,dellanna2006,markos2007,konig2012,mondragon2014,luo2020,li2020,wang2021a,li2022} have garnered renewed interest in recent years~\cite{karcher2023a,karcher2023b,xiao2023,Nayak2024,silva2025,zhao2025} because they share the same universality classes as the phase transitions in the fundamental non-Hermitian symmetry classes~\cite{kawabata2021,luo2022,chen2025}. In addition, it was shown that the Anderson transition in chiral symmetry classes is a topological phase transition driven by the spatial proliferation of vortex excitations~\cite{konig2012,zhao2024,zhao2025} analogous to the two-dimensional (2D) Berezinskii-Kosterlitz-Thouless (BKT) transition~\cite{berezinskii71,berezinskii72,kosterlitz1972,kosterlitz1973,kosterlitz1974} and three-dimensional (3D) superfluid transition~\cite{onsager1949,feynman1955,popov1973,wiegel1973,nelson1977,dasgupta1981,williams1987,shenoy1989,shindou2025}. Thereby, a band topology inherent in underlying lattice models could reshape the delocalization-localization transition in chiral symmetry classes, leading to entirely new disorder-driven phases and phase transitions~\cite{zhao2024,zhao2025,li2025,silva2025,shang2025}. For example, recent studies have revealed that the 1D weak band topology in chiral symmetry classes~\cite{Ryu2010,schnyder2011} universally induces an intermediate quasi-localized phase, bridging metallic and fully localized phases~\cite{xiao2023,zhao2024,silva2025}. In this quasi-localized phase, the exponential localization length diverges exclusively along the spatial direction with the 1D weak topology (the ``topological direction"), while remaining finite in other directions.

Localization phenomena across the Altland-Zirnbauer (AZ) symmetry classes can generally be studied using nonlinear sigma models (NLSMs)~\cite{wegner1976,wegner1979,efetov1980,hikami1980,hikami1981,efetov1983,wegner2016}, where a  matrix field $Q$ resides in a Goldstone manifold constrained by the symmetries of the underlying disordered Hamiltonian~\cite{altland1997,altland1999,fukui1999,fabrizio2000,guruswamy2000}. In chiral symmetry classes, the $Q$ field permits vortex excitations as saddle-point solutions of the NLSM~\cite{konig2012,zhao2024,zhao2025}. Crucially, the 1D weak band topology~\cite{Ryu2010,schnyder2011,fulga2012a} endows these vortex excitations with complex phase factors that depend on their spatial polarization along the topological direction~\cite{altland2014,altland2015,zhao2024}. Previously, it was argued that due to a destructive interference effect caused by these phase factors, a metal phase near its mobility edge could be dominated by vortex excitations polarized along the topological direction, and the spatial proliferation of the polarized vortex excitations may result in the universal emergence of the quasi-localized phase adjacent to the metal phase~\cite{zhao2024,zhao2025,shindou2025}. The intermediate quasi-localized phase has so far been confirmed numerically only in the chiral unitary and orthogonal classes~\cite{xiao2023,zhao2024}, but not yet in the chiral symplectic class.

     In this paper, we investigate the nature of the Anderson transition in chiral symmetry classes with the one-dimensional (1D) weak topology, combining lattice model simulations with a renormalization group (RG) analysis of effective theories. In the numerical study, we extend previous transfer matrix analyses to the chiral symplectic class (CII) and perform a finite-size scaling (FSS) analysis of the localization length. Our simulations confirm that the critical exponent of the Anderson transition in the chiral symplectic class differs markedly from those in the chiral unitary and chiral orthogonal classes, underscoring the significant role of the Kramers time-reversal symmetry in the Anderson transition criticality of the chiral classes. When the 1D weak topology is introduced, the numerics reveal the emergence of an intermediate quasi-localized (QL) phase in 3D and a corresponding QL regime in 2D, where the localization length diverges exclusively along the direction of the weak topology. The estimated critical exponents further demonstrate that the metal-to-QL transition belongs to a universality class distinct from that of the Anderson transition without the 1D topology, and that the critical exponent in the chiral symplectic class differs substantially from those in the other two chiral classes, highlighting the prominent role of the Kramers degeneracy in the metal-to-QL transition criticality.

    Motivated by these numerical findings, we revisit and extend our previous 2D RG analysis to all three chiral classes by including the one-loop renormalization of the weak topological term. The revised analysis shows that, once the one-loop renormalization of the weak topological term is taken into account, the quasi-localized strong-coupling fixed point previously identified in the chiral unitary class becomes unstable and is replaced by a stable fixed point exhibiting conventional localized behavior. Despite this change, the RG analysis continue to capture the influence of the 1D weak topology in both the chiral unitary and chiral symplectic classes, which manifests as spatially anisotropic scaling at the critical point of the Anderson transition. These theoretical results suggest that the quasi-localized regime observed numerically in 2D models may be an artifact of the isotropic scaling ansatz assumed in the FSS analysis. A definitive numerical confirmation of the quasi-localized phase will therefore require a finite-size scaling framework capable of handling generic spatial scaling.

\section{Models}
A Hermitian Hamiltonian with chiral symmetry can be brought into a block off-diagonal structure,
\begin{equation}
    H = \begin{pmatrix}
        0 & h\\
        h^\dagger & 0
    \end{pmatrix}, \label{hermitize}
\end{equation}
where $h$ is generally a non-Hermitian square matrix. For the 3D model in the chiral symplectic class without a weak topological index, the off-diagonal block of the Hamiltonian is given by the following tight-binding model defined on the cubic lattice,
\begin{subequations}
\label{hNT}
\begin{align}
h_{\mathrm{NT}}^{\mathrm{3D}}
= \sum_{\bm{j}} \bigg\{
& i(\Delta+\epsilon_{\bm{j}})
c_{\bm{j}}^\dagger \sigma_z c_{\bm{j}} + iD\,c_{\bm{j}}^\dagger \sigma_x c_{\bm{j}}
+ iF\,c_{\bm{j}}^\dagger \sigma_y c_{\bm{j}}
\nonumber\\
& + t_{||}
c_{\bm{j}+\bm{e}_z}^\dagger \sigma_0 c_{\bm{j}}
- i t_{||}'
c_{\bm{j}+\bm{e}_z}^\dagger \sigma_z c_{\bm{j}}
\nonumber\\
& - t_{||}
c_{\bm{j}-\bm{e}_z}^\dagger \sigma_0 c_{\bm{j}}
+ i t_{||}'
c_{\bm{j}-\bm{e}_z}^\dagger \sigma_z c_{\bm{j}}
\nonumber\\
& + i t_\perp
c_{\bm{j}+\bm{e}_x}^\dagger \sigma_x c_{\bm{j}}
- i t_\perp
c_{\bm{j}-\bm{e}_x}^\dagger \sigma_x c_{\bm{j}}
\nonumber\\
& - i t_\perp
c_{\bm{j}+\bm{e}_y}^\dagger \sigma_y c_{\bm{j}}
+ i t_\perp
c_{\bm{j}-\bm{e}_y}^\dagger \sigma_y c_{\bm{j}}
\bigg\}.
\label{hNT_3D}
\end{align}

For the 2D model, we denote the two spatial directions by $x$ and $y$
and define the off-diagonal block on the square lattice as
\begin{align}
h_{\mathrm{NT}}^{\mathrm{2D}}
= \sum_{\bm{j}} \bigg\{
& i(\Delta+\epsilon_{\bm{j}})
c_{\bm{j}}^\dagger \sigma_z c_{\bm{j}}+ iD\,c_{\bm{j}}^\dagger \sigma_x c_{\bm{j}}
+ iF\,c_{\bm{j}}^\dagger \sigma_y c_{\bm{j}}
\nonumber\\
& + t_{||}
c_{\bm{j}+\bm{e}_y}^\dagger \sigma_0 c_{\bm{j}}
- i t_{||}'
c_{\bm{j}+\bm{e}_y}^\dagger \sigma_z c_{\bm{j}}
\nonumber\\
& - t_{||}
c_{\bm{j}-\bm{e}_y}^\dagger \sigma_0 c_{\bm{j}}
+ i t_{||}'
c_{\bm{j}-\bm{e}_y}^\dagger \sigma_z c_{\bm{j}}
\nonumber\\
& - i t_\perp
c_{\bm{j}+\bm{e}_x}^\dagger \sigma_y c_{\bm{j}}
+ i t_\perp
c_{\bm{j}-\bm{e}_x}^\dagger \sigma_y c_{\bm{j}}
\bigg\}.
\label{hNT_2D}
\end{align}
\end{subequations}
Here the subscript ``$\mathrm{NT}$'' denotes the ``non-topological'', in contrast to the subscript ``$\mathrm{T}$'' (topological) introduced later in the Hamiltonian. For the 3D cubic-lattice model, $c_{\bm{j}}$ is a two-component annihilation operator defined on the cubic lattice site $\bm{j}=(j_x,j_y,j_z)$ with $j_{x,y,z}\in \mathbb{Z}$, and ${\bm e}_{\mu}$ is the unit vector along the direction $\mu\in\{x,y,z\}$. For the 2D square-lattice model, $\bm{j}=(j_x,j_y)$ and $\mu\in\{x,y\}$.  $\sigma_0$ and $\sigma_{\mu=x,y,z}$ are the 2-by-2 unit matrix and Pauli matrices, respectively. $\Delta, D, F, t_{||}, t_{||}'$, and $t_{\perp}$ are real-valued parameters, and $\epsilon_{\bm{j}}$ is a random potential distributed uniformly in the range $[-W/2,W/2]$ with a disorder strength $W\ge0$. The non-Hermitian Hamiltonian respects the
time-reversal symmetry $\sigma_y(h_{\mathrm{NT}}^{d})^*\sigma_y^\dagger
=h_{\mathrm{NT}}^{d}$, where $d=\mathrm{2D},\mathrm{3D}$, so that the corresponding Hermitian Hamiltonian obtained from Eq.~\eqref{hermitize} belongs to the chiral symplectic class.

For the 3D model with a weak topological index, the off-diagonal block of the Hamiltonian is given by the following tight-binding model on the cubic lattice,

\begin{subequations}
\label{hT}
\begin{align}
h_{\mathrm{T}}^{\mathrm{3D}}
= \sum_{\bm{j}} \bigg\{
& i\Delta\,
c_{\bm{j}}^\dagger \sigma_x c_{\bm{j}}
+ i\epsilon_{\bm{j}}\,
c_{\bm{j}}^\dagger \sigma_z c_{\bm{j}}+ i\epsilon_{\bm{j}}'\,
c_{\bm{j}}^\dagger \sigma_y c_{\bm{j}}
\nonumber\\
& + i(t_{||}+t_{||}')
c_{\bm{j}+\bm{e}_z}^\dagger
\sigma_x c_{\bm{j}}+ i(t_{||}-t_{||}')
c_{\bm{j}-\bm{e}_z}^\dagger
\sigma_x c_{\bm{j}}
\nonumber\\
& + i t_\perp
\sum_{\mu=x,y}
\left(
c_{\bm{j}+\bm{e}_\mu}^\dagger
\sigma_x c_{\bm{j}}
+
c_{\bm{j}-\bm{e}_\mu}^\dagger
\sigma_x c_{\bm{j}}
\right)
\bigg\}.
\label{hT_3D}
\end{align}

For the 2D model, the off-diagonal block on the square lattice is
\begin{align}
h_{\mathrm{T}}^{\mathrm{2D}}
= \sum_{\bm{j}} \bigg\{
& i\Delta\,
c_{\bm{j}}^\dagger \sigma_x c_{\bm{j}}
+ i\epsilon_{\bm{j}}\,
c_{\bm{j}}^\dagger \sigma_z c_{\bm{j}}+ i\epsilon_{\bm{j}}'\,
c_{\bm{j}}^\dagger \sigma_y c_{\bm{j}}
\nonumber\\
& + i(t_{||}+t_{||}')
c_{\bm{j}+\bm{e}_y}^\dagger
\sigma_x c_{\bm{j}}+ i(t_{||}-t_{||}')
c_{\bm{j}-\bm{e}_y}^\dagger
\sigma_x c_{\bm{j}}
\nonumber\\
& + i t_\perp
c_{\bm{j}+\bm{e}_x}^\dagger
\sigma_x c_{\bm{j}}+ i t_\perp
c_{\bm{j}-\bm{e}_x}^\dagger
\sigma_x c_{\bm{j}}
\bigg\}.
\label{hT_2D}
\end{align}
\end{subequations}
Here, $\Delta,t_{||},t_{||}'$, and $t_{\perp}$ are real-valued parameters, and $\epsilon_j, \epsilon_j'$ are independent random potentials distributed uniformly in the range of $[-W/2,W/2]$. Both $h_{\mathrm{T}}^{\mathrm{3D}}$ and $h_{\mathrm{T}}^{\mathrm{2D}}$ have only the time-reversal symmetry $\sigma_y(h_{\mathrm{T}}^{d})^*\sigma_y^\dagger=h_{\mathrm{T}}^{d}$, where $d=\mathrm{2D},\mathrm{3D}$; the corresponding Hermitian Hamiltonians belong to the chiral symplectic class. Note that $t_{\parallel}t_{\parallel}'\neq0$ introduces nonreciprocal hopping along the $z$ direction in 3D and along the $y$ direction in 2D, giving rise to the weak topology along the respective direction (see Sec.~\ref{sec3c})~\cite{esaki2011,yao2018,kawabata2019a}.

\section{Methods}

\subsection{Transfer Matrix Method}
The universality class of continuous phase transitions is characterized by a critical exponent $\nu$ associated with the divergent length scale $\xi$ near a critical point, $\xi \propto |W-W_c|^{-\nu}$. To evaluate the critical exponent of the disorder-driven quantum phase transitions, we consider the transmission probability of plane waves through the disordered system with a quasi-one-dimensional (Q1D) geometry. The linear dimension $L_{\mu}$ along the spatial direction of transmission $\mu$ is much larger than the linear dimension $L$ of the other dimensions, i.e., $L \ll L_{\mu}$~\cite{mackinnon1981,pichard1981,mackinnon1983}. We calculate the longest exponential decay length $\xi_{\mu}$ of the transmission probability along $\mu$ direction. The transmission is calculated from a sequential product of the transfer matrices $M$ of the disordered system along $\mu$ direction. Since the transmission probability in Q1D disordered systems for finite $L$ generally decays exponentially in $L_{\mu}$, all eigenvalues of the following matrix take finite real values,
\begin{equation}
    \lim_{L_\mu \rightarrow \infty}\log(M^\dagger M)^{\frac{1}{2L_\mu}}.
\end{equation}
These eigenvalues are called Lyapunov exponents (LEs) of the disordered $H$. Positive and negative LEs calculated from the transfer matrix product along the $+\mu$ direction represent the inverse of exponential decay lengths of the transmission along $+\mu$ and $-\mu$ direction, respectively.  Due to the hermiticity of $H$, a set of the LEs of $H$ is symmetric with respect to the sign change. Thus, the longest decay length $\xi_{\mu}\equiv \gamma^{-1}_{\mu}$ is given by the smallest positive LE. Notably, due to the chiral symmetry, the LEs of $H$ in Eq.~\eqref{hermitize} are equal to a sum of the LEs of $h$ and those of $h^\dagger$, where the LEs of $h$ and those of $h^{\dagger}$ come in the pairs with opposite signs to each other~\cite{crisanti2012,xiao2023}. To analyze $\xi_{\mu}$, we calculate the LEs of $h$ from the transfer-matrix product $m$ of disordered $h$ along $\mu$~\cite{luo2021b}.

\subsection{Distribution of the LEs of $h$}
An ensemble of disordered Hamiltonians can be invariant under certain symmetry transformations. Such statistical symmetry~\cite{fulga2012a} constrains the distributions of the LEs in the thermodynamic limit $(L_{\mu} \rightarrow \infty)$. The ensemble of $h_{\rm NT}$ is statistically symmetric under a spatial inversion
\begin{align}
    &\{h_{\mathrm{NT}} \, | \, \epsilon_j \in [-W/2, W/2]\}  \nonumber \\
    & =\{\mathcal{U}_1 \!\ h_{\mathrm{NT}} \!\ \mathcal{U}^{\dagger}_1 \,| \, \epsilon_j \in [-W/2,W/2]\},
    \label{statissymm_nt}
\end{align}
with $(\mathcal{U}_1)_{{\bm j},{\bm j}^{\prime}} = (-1)^{j_x+j_y+j_z} \delta_{{\bm j},-{\bm j}^{\prime}}$ for the 3D $h_{\rm NT}$ model and  $(\mathcal{U}_1)_{{\bm j},{\bm j}^{\prime}} = (-1)^{j_x+j_y} \delta_{{\bm j},-{\bm j}^{\prime}}$ for the 2D $h_{\rm NT}$ model. Since the inversion flips the sign of the LEs along all spatial directions, Eq.~\eqref{statissymm_nt} means that the LEs of $h_{\rm NT}$ must come in opposite-sign pairs for any spatial direction $\mu$ ~\cite{xiao2023,zhao2024}.

\begin{figure*}
    \centering
    \includegraphics[width=0.9\linewidth]{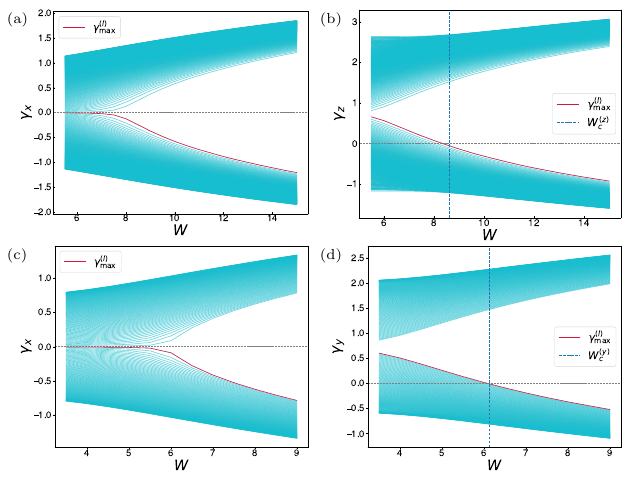}
    \caption{\label{LEs_3DL16_2DL120} Distributions of the Lyapunov exponents (LEs) of 3D $h_{\mathrm{T}}$ along non-topological $x$ (a) and topological $z$ directions (b), and of the 2D $h_{\mathrm{T}}$ model along non-topological $x$ (c) and topological $y$ directions (d), as a function of disorder strength $W$. Parameters in Eq.~\eqref{hT} are set as $\Delta = 0.1, t_{||} = 0.5, t_{||}' = 0.8, t_{\perp} = 1$. System sizes along the transverse directions are $L^2 = 256$ for 3D and $L = 120$ for 2D. With moderately larger $W$, the LEs comprise two separate bands, each containing $2L^2$ (3D) or $2L$ (2D) LEs. $\gamma_{\mathrm{max}}^{(l)}$ is the largest LE of the lower LE band. For the non-topological directions shown in (a) and (c), the LE distributions are symmetric about zero. For the topological directions shown in (b) and (d), the lower LE band covers zero in the smaller-$W$ region, where the localization length along the topological direction is divergent and the corresponding weak topological index takes a non-quantized value. Blue dashed lines in (b) and (d) mark $W_c^z$ and $W_c^y$, respectively, determined by the finite-size scaling in Eq.~\eqref{scaling_tp}.}
\end{figure*}

The ensemble of $h_{\rm T}$ is statistically symmetric under a mirror reflection about the plane perpendicular to the topological direction followed by a transposition,
\begin{align}
    &\{h_{\mathrm{T}} \, | \,\epsilon_j, \epsilon_j' \in [-W/2, W/2]\} \nonumber \\
    &=\{\mathcal{U}_2 \!\  h^T_{\mathrm{T}} \!\ \mathcal{U}^{\dagger}_2\,| \, \epsilon_j, \epsilon_j' \in [-W/2,W/2]\}, \label{statissymm_t}
\end{align}
with $(\mathcal{U}_2)_{{\bm j},{\bm j}^{\prime}} =  \delta_{j_x,j^{\prime}_x} \delta_{j_y,j^{\prime}_y}\delta_{j_z,-j^{\prime}_z}$ for the 3D $h_{\rm T}$ model and $(\mathcal{U}_2)_{{\bm j},{\bm j}^{\prime}} =  \delta_{j_x,j^{\prime}_x}\delta_{j_y,-j^{\prime}_y}$ for the 2D $h_{\rm T}$ model. Since the transposition also reverses directions of the transmission, the mirror reflection followed by the transposition changes the sign of the LEs along the non-topological directions, while it does not influence the LEs along the topological direction. Therefore, in the thermodynamic limit, the LEs of the 3D $h_{\mathrm{T}}$ model along the $x$ and $y$ directions and those of the 2D $h_{\mathrm{T}}$ model along the $x$ direction must come in opposite-sign pairs. The LEs along the topological direction, namely $z$ in 3D and $y$ in 2D, are not constrained by this symmetry
(see Fig.~\ref{LEs_3DL16_2DL120}). The asymmetry becomes more prominent in a weakly disordered region, where the numbers of positive and negative LEs along the topological direction are different. Their imbalance is directly related to the corresponding weak topological index~\cite{luca2003,fulga2011,xiao2023}.

\subsection{1D weak topology}
\label{sec3c}
The 1D weak topological index in the chiral symmetry class is defined by the off-diagonal non-Hermitian matrix $h$ with a twisted boundary condition: a particle traversing around the torus of $h(\phi_{\mu})$ along $\mu$ direction acquires a U(1) phase $e^{i\phi_{\mu}}$~\cite{Ryu2010,fulga2011,mondragon2014,altland2014,claes2020}. It counts how many times $\det[h(\phi_{\mu})]$ winds about the origin in the complex plane when $\phi_{\mu}$ is changed from $0$ to $2\pi$,
\begin{align}
    \nu_{\mu} = \frac{i}{\cal N}\int_0^{2\pi}\frac{d\phi_{\mu}}{{2\pi}}\partial_{\phi_{\mu}} \log \{\det[h(\phi_{\mu})]\}.
\end{align}
Here, the winding number is normalized by ${\cal N}=2L^2$ in 3D and ${\cal N}=2L$ in 2D for the models defined in Eqs.~\eqref{hNT} and \eqref{hT}. The topological index thus defined is quantized to integer values in a gapped (localized) phase, while it is not quantized in a metal phase. Importantly, the weak topological index $\nu_{\mu}$ is related to the imbalance between the number of positive and negative LEs of $h \equiv h(\phi_{\mu}=0)$ along $\mu$~\cite{luca2003,fulga2011,xiao2023},
\begin{align}
    \nu_{\mu} \equiv \frac{N_{+,\mu}-N_{-,\mu}}{N_{+,\mu}+N_{-,\mu}}.
\end{align}
Here, $N_{\pm,\mu}$ is the total number of positive (negative) LEs of $h$ along $\mu$ direction. For $h_{\rm NT}$ in Eq.~\eqref{hNT}, the statistical symmetry in Eq.~\eqref{statissymm_nt} ensures $N_{+,\mu}=N_{-,\mu}$ for all directions $\mu$ so that $\nu_{\mu}=0$. For $h_{\rm T}$ in Eq.~\eqref{hT}, the statistical symmetry in Eq.~\eqref{statissymm_t} ensures $\nu_x=\nu_y=0$, while $\nu_z$ can be finite in 3D; in 2D, it ensures $\nu_x=0$, while $\nu_y$ can be finite.

In the presence of moderately strong disorder, the LEs of the 3D and 2D models defined in Eqs.~\eqref{hNT} and \eqref{hT} form two separate bands, where each band contains $2L^2$ and $2L$ LEs (see Fig.~\ref{LEs_3DL16_2DL120}). For the LEs of $h_{\rm T}$ along the topological direction, one of the two LE bands covers the zero in a weakly disordered regime [see Fig.~\ref{LEs_3DL16_2DL120}(b,d)], where the Q1D localization length along the topological direction is divergent and the corresponding weak topological index takes a non-integer finite value in the limit of $L\rightarrow \infty$. The boundary of such a metallic phase can be determined by a finite-size scaling analysis of the LEs on the band edge.

The scaling analysis of the band edge assumes a finite gap $\Delta(W,L)$ between the largest LE $\gamma^{(l)}_{\rm max}(W,L)$ of the lower LE band and the smallest LE $\gamma_{\rm min}^{(u)}(W,L)$ of the upper LE band. When the gap $\Delta$ is much larger than $L^{-1}$, $\gamma_{\mathrm{max}}^{(l)}(W,L)$ follows the scaling function~\cite{hatano1996,asada2004,xiao2023}
\begin{equation}
    \gamma_{\mathrm{max}}^{(l)}(W,L) = -\frac{a}{L} + \gamma_{\mathrm{max}}^{(l)}(W).
    \label{scaling_tp}
\end{equation}
Using a standard linear regression, we determine $\gamma_{\mathrm{max}}^{(l)}(W) \equiv \lim_{L\rightarrow \infty}\gamma^{(l)}_{\rm max}(W,L)$. For the LEs of $h_{\rm T}$ along the topological direction $\mu_{\mathrm{t}}$, where $\mu_{\mathrm{t}}=z$ in 3D and $\mu_{\mathrm{t}}=y$ in 2D, $\gamma^{(l)}_{\rm max}(W=W_c^{\mu_{\mathrm{t}}})=0$ defines the boundary $W_c^{\mu_{\mathrm{t}}}$ of the metallic region with divergent $\xi_{\mu_{\mathrm{t}}}$ and finite $\nu_{\mu_{\mathrm{t}}}$. When $\gamma^{(l)}_{\rm max}(W)$ crosses zero linearly at $W=W_c^{\mu_{\mathrm{t}}}$, the critical exponent at the phase transition is $1$.

For the LEs of $h_{\rm T}$ along the non-topological directions, namely $x$ and $y$ in 3D and $x$ in 2D, the two LE bands come in the opposite-sign pairs as shown in Fig.~\ref{LEs_3DL16_2DL120}(a,c). Likewise, the two LE bands of $h_{\mathrm{NT}}$ along all directions come in the opposite-sign pairs similar to Fig.~\ref{LEs_3DL16_2DL120}(a,c). Thereby, $\xi_\mu^{-1}$ does not cross zero at finite $L$ for $\mu=x,y$ in the 3D $h_{\mathrm{T}}$ model, for $\mu=x$ in the 2D $h_{\mathrm{T}}$ model, or for any spatial direction in $h_{\mathrm{NT}}$, while $\xi_\mu$ still undergoes a delocalization-localization transition in the thermodynamic limit ($L\to\infty$). To study the delocalization-localization transition along these non-topological directions, we introduce a more general finite-size-scaling form of $\xi_\mu(W,L)$ in the next section.

\subsection{Finite-Size Scaling}
\label{sec3d}
The criticality of the phase transition is typically determined by a saddle fixed point of some renormalization group equations. Here, we assume that at this postulated saddle-fixed point, a length scale along all spatial directions scales with the same exponent, corresponding to spatially isotropic scaling. Then, the standard scaling theory dictates that a normalized length scale, $\Lambda_{\mu} \equiv \xi_{\mu}/L$, should be a function only of $\phi_t \equiv L^{y_t} u_t(w)$, and $\phi_i \equiv L^{-y_i} u_i(w)$ ($i=1,2,\cdots$), where $u_t(w)$ is a relevant scaling variable around the saddle fixed point with its scaling dimension $y_t>0$, and $u_i(w)$ are irrelevant scaling variables with their negative scaling dimensions $-y_i$, and $0<y_1<y_2<\cdots$. These scaling variables are functions of normalized disorder strength $w\equiv (W-W_c)/W_c$.  In the analysis, we keep only the relevant scaling variable $u_t(w)$ and the least irrelevant scaling variable $u_1(w)$, and expand a scaling function with respect to $\phi_t$ and $\phi_1 $~\cite{slevin1999,slevin2014};
\begin{align}
    \ln \Lambda_{\mu}(W,L) &= F(L^{y_t} u_t, L^{-y_1} u_1)  \nonumber \\
    &= \sum_{i=0}^{n_1} \sum_{j=0}^{n_2}a_{ij}(L^{1/\nu}u_t)^{i}(L^{-y_1} u_1)^{j},
    \label{scaling}
\end{align}
with the critical exponent $\nu \equiv 1/y_t$. For the sake of stable fitting, data of $\ln\Lambda_{\mu}$ rather than $\Lambda_{\mu}$ are studied using finite-size scaling analysis. Near the critical point $W_c$, the two scaling variables can be further expanded in a series of $w$,
\begin{equation}
    u_t = \sum_{k = 1}^{m_1}b_kw^k, \quad u_1 = \sum_{k = 0}^{m_2}c_kw^k.\label{scaling2}
\end{equation}
To determine the critical exponent $\nu$ as well as  the universal single-parameter scaling function $F(\phi_t,0)$ for small $\phi_t$, we fit the data points using Eqs.~(\ref{scaling},\ref{scaling2}) with finite orders $n_1$, $n_2$, $m_1$, and $m_2$ of the truncation. The fitting is performed by minimizing the following chi-square function with respect to polynomial fitting parameters such as $\nu$, $y_1$, $W_c$, $a_{ij},$ ${b_k}$ and $c_k$,
\begin{equation}
    \chi^2 = \sum_{i = 1}^{N_D}\frac{(\ln \Lambda_i - F_i)^2}{\sigma_i^2}. \label{chi2}
\end{equation}
Here, $\ln \Lambda_i$ and $\sigma_i$ are the mean value and standard deviation of $\ln \Lambda(W,L)$ for the $i$th data point, respectively. $F_i$ is a value from the polynomial fitting function for the same data point. The total number of data points $N_D$ is larger than \begin{revision}180\end{revision} in all the polynomial FSS fittings in this paper. Without loss of generality, we set $a_{10}=a_{01}=1$, by redefining $a_{10} b_{k}$, $a_{01} c_{k}$ and $a_{ij}/(a^i_{10} a^j_{01})$ ($i+j\ge 2$) as new $b_{k}$,  $c_k$ and $a_{ij}$, respectively.

\section{Results}

\subsection{\label{subsec:Nontopological_Model_in_3D}3D model without 1D weak topology}

\begin{figure*}
    \centering
    \includegraphics[width=0.9\linewidth]{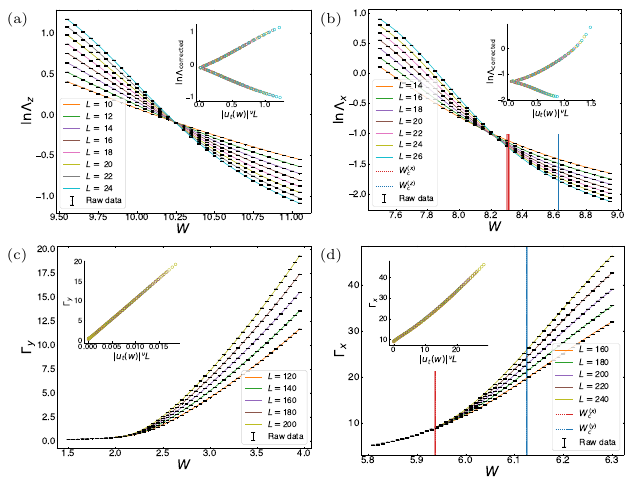}
    \caption{\label{FSS_figure} Logarithm (inverse) of the normalized localization length $\ln\Lambda = \ln \xi/L$ ($\Gamma = 1/\Lambda$) as a function of the disorder strength $W$ for different system sizes $L$. (a) The length $\xi_z$ along $z$ in 3D $h_{\rm NT}$ (non-topological model); (b) the length $\xi_x$ along the non-topological $x$ direction in 3D $h_{\rm T}$ (topological model); (c) the length $\xi_y$ along $y$ in 2D $h_{\rm NT}$ (non-topological model); (d) the length $\xi_x$ along the non-topological $x$ direction in 2D $h_{\rm T}$ (topological model). Black dots with error bars are numerical data from the transfer matrix calculation, where the error bars are around 0.5\%, except for the error bars for the data points from a weakly disordered regime in (c). To ensure a reasonable accuracy in (c), we choose $L_y = 5\times 10^6$. Solid lines in (a) and (b) are fitting curves by a scaling function of Eq.~\eqref{scaling}, whose parameters are from the fitting result with the largest goodness of fit (GOF) [see Table~\ref{FSStable_3D}]. Solid lines in (c) and (d) are fitting curves by a single-parameter scaling function of Eq.~\eqref{scaling_2D}, whose parameters are from the fitting results with $n =2, m = 3$ and $n = m = 2$, respectively. The red vertical lines in (b) and (d) indicate locations of the critical disorder strengths $W_c^x$ estimated by the fitting. The blue vertical lines in (b) and (d) indicate $W_c^z$ for the 3D model and $W_c^y$ for the 2D model, respectively; below these values, the localization length along the topological direction is divergent and the corresponding weak topological index takes non-quantized values. Shaded regions around the vertical lines denote the $95\%$ confidence intervals obtained from Monte Carlo simulations. Insets of each panel show the (corrected) single-parameter scaling functions as a function of small $|\phi_t|^{\nu} \equiv |u_t|^{\nu}L$.}
\end{figure*}

We first study the Anderson transition of the zero-energy eigenstates of the 3D model without a weak topological index in Eq.~\eqref{hNT_3D}, with a set of parameters $\Delta = D = F = t_{\perp} = t_{||} = 1$, $t_{||}' = 0.5$. The Q1D localization length $\xi_{z}$ of different cross-sectional sizes $L$ as a function of disorder strength $W$ is shown in Fig.~\ref{FSS_figure}(a). In the clean limit, the system is in a semimetal phase (see Fig.~\ref{DOS_CII_3D_wo} for $W=0$). The critical point of an Anderson transition is manifested by the scale-invariant point of a normalized localization length $\Lambda_z = \xi_z/L$. We analyze the data points of $\ln \Lambda_z$ obtained by the transfer matrix method near the critical point with the polynomial fitting function introduced in Sec.~\ref{sec3d}.

The fitting results are summarized in Table~\ref{FSStable_3D}. To test the stability of the polynomial fitting, we increase the truncation orders $n_1,m_1$ associated with the relevant scaling variable, while keeping the truncation order related to the least irrelevant scaling variable minimal, i.e., $n_2=1$ and $m_2=0,1$. Our fitting results with the best goodness of fit (GOF) show that the critical exponent of the Anderson transition in the 3D chiral symplectic class with the estimated standard error is \begin{revision}$\nu_{\rm CII} = 0.868\pm 0.006$\end{revision}. Notably, the zero-energy DOS remains finite at the estimated critical point [see Fig.~\ref{DOS_CII_3D_wo}(b)]. The critical exponent is remarkably consistent with a previous study of the Anderson transition in the non-Hermitian symmetry class AII~\cite{luo2022}, providing further proof of the correspondence of universality classes of Anderson transitions between Hermitian and non-Hermitian systems~\cite{kawabata2021,luo2022,chen2025}. The critical exponent of the 3D chiral symplectic class is also significantly different from previously reported critical exponents in the chiral unitary class $\nu_{\mathrm{AIII}} = 1.06\pm 0.02$ and chiral orthogonal class $\nu_{\mathrm{BDI}} = 1.12\pm 0.06$~\cite{wang2021a,luo2022,xiao2023}, highlighting the importance of Kramers time-reversal symmetry in determining the critical behaviors of the Anderson transition in chiral classes. Combining the optimal fitting parameters and numerical data, the single-parameter scaling function $F(\phi_t,0)$ for small $\phi_t$ is obtained from $\ln\Lambda_{\rm corrected}(W,L) \equiv \ln \Lambda(W,L) - \sum^{n_1}_{i=0}\sum^{n_2}_{j=1} a_{ij}(\phi_t)^i(\phi_1)^j$ [Fig.~\ref{FSS_figure}(a)]. When plotted as a function of $|\phi_t(W,L)|^{\nu}\equiv L|u_t(w)|^{\nu}$, all the corrected data points from different $W$ and $L$ collapse into a single-parameter scaling function. The upper and lower branches in Fig.~\ref{FSS_figure}(a) correspond to the single-parameter scaling function in the metal and insulator phase sides, respectively.

\subsection{\label{subsec:Topological_model_in_3D}3D model with 1D weak topology}

We next study the disorder-driven quantum phase transitions of the zero-energy eigenstates of the 3D model with a weak topological index in Eq.~\eqref{hT_3D}, with a set of parameters $\Delta = 0.1, t_{||} = 0.5, t_{||}' = 0.8, t_{\perp} = 1$. The hopping amplitudes are reciprocal in the $x$ and $y$ directions, while $t_{\parallel}t^{\prime}_{\parallel} \ne 0$ breaks the reciprocity along the $z$ direction. The (non)reciprocity of $h_{\rm T}$ results in the (a)symmetric distribution of the LEs about zero. Fig.~\ref{LEs_3DL16_2DL120}(a) and (b) show the distribution of the LEs along the $x$ and $z$ directions, respectively. The asymmetric distribution of the LEs along $z$ becomes more prominent for $W<W_c^z$, where the total number $N_{+,z}$ of the positive LEs differs from the total number $N_{-,z}$ of the negative LEs. For $W<W_c^z$, the lower LE band covers zero, resulting in divergent $\xi_z$ and non-quantized finite $\nu_z$. The boundary $W_c^z$ of the metallic region with divergent $\xi_z$ and non-zero $\nu_z$ is determined by the linear regression of $\gamma^{(l)}_{\rm max}(W,L)$ [Sec.~\ref{sec3c}]. We obtain \begin{revision}$W_c^z=8.6250\pm 0.0009$ \end{revision}and confirm that $\gamma^{(l)}_{\rm max}(W)$ crosses zero linearly at $W=W_c^z$. Therefore, the critical exponent of this transition is $\nu=1$.

\begin{table*}[htb]
    \caption{\label{FSStable_3D} Finite-size scaling (FSS) analyses in the 3D models with and without 1D topology. (a) FSS analyses of the normalized localization length of 3D $h_{\rm NT}$ along $z$; (b) FSS analyses of the normalized localization length of 3D $h_{\rm T}$ along the non-topological $x$ direction. $n_1$, $m_1$, $n_2$ and $m_2$ are truncation orders of the polynomial expansion of the scaling function [Eq.~\eqref{scaling} and Eq.~\eqref{scaling2}]. $L$ denotes a data range of the system size used for the FSS analyses. $W_c$ [$W_c^x$ in (b)], $\nu$, $-y_1$ and $\Lambda_c$ are the critical disorder strength, critical exponent, scaling dimension of the least irrelevant scaling variable, and value of the normalized length $\Lambda$ at the critical point [determined by the $\chi^2$ minimization]. The square brackets in each item indicate 95\% confidence intervals [determined from Monte Carlo simulations of 1000 samplings]. }
    \begin{ruledtabular}
    \begin{tabular}{cccccccccc}
        \multicolumn{5}{l}{(a) 3D without topology}\\[0.1cm]
        $n_1$&$m_1$&$n_2$&$m_2$&$L$&GOF&$W_c$&$\nu$&$y_1$&$\Lambda_c$\\
        3&3&1&1&10--22
        &0.460
        &10.249[10.244,10.255]
        &0.883[0.866,0.906]
        &1.160[0.618,1.496]
        &0.906[0.893,0.913]\\
        4&3&1&1&10--22
        &0.687
        &10.244[10.239,10.250]
        &0.873[0.859,0.888]
        &1.676[1.118,4.711]
        &0.914[0.905,0.920]\\
        3&3&1&1&10--24
        &0.300
        &10.249[10.245,10.253]
        &0.881[0.867,0.898]
        &1.044[0.719,1.252]
        &0.905[0.897,0.912]\\
        4&3&1&1&10--24
        &0.810
        &10.243[10.240,10.247]
        &0.868[0.857,0.880]
        &1.587[1.063,2.761]
        &0.915[0.909,0.920]\\
        3&2&1&1&10--24
        &0.214
        &10.249[10.244,10.253]
        &0.887[0.867,0.907]
        &0.618[0.434,0.896]
        &0.902[0.891,0.913]\\
        3&3&1&0&10--24
        &0.0054
        &10.244[10.241,10.245]
        &0.862[0.852,0.872]
        &1.806[1.662,1.942]
        &0.914[0.913,0.919]\\[0.5cm]

        \multicolumn{8}{l}{(b) 3D with topology (non-topological direction)}\\[0.1cm]
        $n_1$&$m_1$&$n_2$&$m_2$&$L$&GOF&$W_c^x$&$\nu$&$y_1$&$\Lambda_c$\\
        4&3&1&1&12--24
        &0.184
        &8.292[8.284,8.300]
        &0.597[0.548,0.632]
        &1.338[1.222,1.459]
        &0.278[0.270,0.286]\\
        4&4&1&1&12--24
        &0.173
        &8.292[8.280,8.299]
        &0.605[0.543,0.699]
        &1.352[1.242,1.627]
        &0.279[0.270,0.293]\\
        4&3&1&1&14--26
        &0.264
        &8.312[8.306,8.320]
        &0.613[0.575,0.638]
        &1.421[1.311,1.537]
        &0.267[0.261,0.274]\\
        4&4&1&1&14--26
        &0.308
        &8.310[8.300,8.318]
        &0.642[0.587,0.697]
        &1.492[1.341,1.789]
        &0.271[0.262,0.283]\\
        4&2&1&1&12--24
        &0.054
        &8.298[8.290,8.308]
        &0.597[0.569,0.629]
        &1.210[1.077,1.334]
        &0.270[0.258,0.278]
    \end{tabular}
    \end{ruledtabular}
\end{table*}

Importantly, the Q1D localization length $\xi_{x}$ along the non-topological $x$ direction undergoes another delocalization-localization transition within the metallic region, with divergent $\xi_z$ and non-quantized finite $\nu_z$. The criticality of the transition is analyzed by the finite-size scaling analysis of $\ln \Lambda_x \equiv \ln \xi_x-\ln L$ introduced in Sec.~\ref{sec3d}. Fitting results are summarized in Table~\ref{FSStable_3D}. The fitting with the best GOF gives \begin{revision}$W_c^x =8.310\pm 0.005$\end{revision} and the critical exponent \begin{revision}$\nu = 0.64\pm 0.03$\end{revision}. Notably, the standard errors of $W_c^x$ and $W_c^z$ are at most less than 2\% of the distance between the two mean values, indicating that an intermediate quasi-localized (QL) phase with divergent $\xi_z$ and finite $\xi_x=\xi_y$ emerges between conventional metal and localized phases. In the metal phase $(W < W_c^x)$, the localization length is divergent along all directions. In the intermediate QL phase $(W_c^x<W<W_c^z)$, the localization length is divergent only along the topological $z$ direction, while it is finite along the non-topological $x$ and $y$ directions. In the localized phase $(W>W_c^z)$, the localization length is finite along all spatial directions~\cite{xiao2023}.

\begin{figure}[htb]
    \centering
    \includegraphics[width=1.0\linewidth]{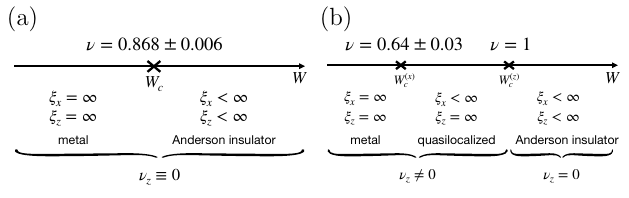}
    \caption{\label{pd_3D} Schematic phase diagrams of 3D models in the chiral symplectic class: (a) without and (b) with 1D weak topology, including critical exponents at their respective phase transition points.}
\end{figure}

The critical exponent \begin{revision}$\nu=0.64 \pm 0.03$ \end{revision}of the metal-to-QL transition is significantly different from the exponent \begin{revision}$\nu = 0.868 \pm 0.006$\end{revision} of the Anderson transition (without the 1D weak topology). This indicates that the 1D weak topology induces a new universality class of the localization-delocalization transition in chiral symmetry classes. Notably, the critical exponent \begin{revision}$\nu_{\mathrm{CII}} = 0.64\pm 0.03$\end{revision} of the metal-to-QL transition deviates largely from the exponents of the metal-to-QL transition in other chiral symmetry classes, $\nu_{\mathrm{AIII}}=\nu_{\mathrm{BDI}} = 0.82\pm 0.02$, highlighting the unique role of the Kramers time-reversal symmetry in the criticality of the metal-to-QL transition~\cite{xiao2023}.

\subsection{2D models}

The Anderson transition is studied for zero-energy eigenstates of 2D $h_{\rm NT}$ in Eq.~\eqref{hNT_2D} with $\Delta  = t_\perp = 1$, $t_{\parallel}=1.25$. The on-site $D, F$  and hopping $t_{||}'$ as well as $\epsilon_{\bm j}$ are treated as local random numbers distributed uniformly and independently in the range $[-W/2,W/2]$, i.e., $D$, $F$, $t^{\prime}_{\parallel} \rightarrow D_{\bm j}, F_{\bm j}, t^{\prime}_{\parallel,{\bm j}} \in [-W/2,W/2]$. The zero-energy eigenstates in the clean limit $(W=0)$ have a higher symmetry than those with disorder $W$, belonging to the Wigner-Dyson (WD) symplectic class. Normalized localization length $\Gamma^{-1}_{y} = \xi_{y}/L$ along $y$ is calculated as a function of $W$ and $L$ using the transfer matrix method. As shown in Fig.~\ref{FSS_figure}(c),  $\Gamma_y(W,L)$ in small and large disorder $W$ regions exhibits scale-invariant and localization behaviors, respectively, indicating the presence of the 2D Anderson transition in the chiral symplectic class. Notably, the scale-invariant behavior in the small $W$ region suggests a 2D critical phase rather than the 2D metal phase in the WD symplectic class~\cite{asada2002}.  A similar critical-metal behavior was also reported in the 2D chiral unitary class~\cite{zhao2024}.

The transition point $W_c$ and critical exponent $\nu$ of the 2D Anderson transition are determined by the polynomial expansion of the single-parameter scaling function~\cite{mackinnon1981,mackinnon1983,slevin2014},
\begin{equation}
    \Gamma_y(W,L) = \sum_{i = 0}^{n}a_i(L^{1/\nu}u_t)^i, \quad u_t = \sum_{k=1}^{m}b_kw^k,
    \label{scaling_2D}
\end{equation}
with $w \equiv (W-W_c)/W_c$. Since $\Gamma_y(W,L)$ in the critical-metal side $W<W_c$  is scale invariant, only data points of $\Gamma_y(W,L)$  from the localized-phase side $W>W_c$ are fitted by the scaling function. We minimize $\chi^2$ iteratively, updating $W_c$ and retaining only data points with $W>W_c$ for the subsequent fit, until $W_c$ converges~\cite{zhao2024}. We examine the results across different initial guesses and values of $n$ and $m$. A stable fitting result is obtained for $n=2$ upon increasing $m$ to 3 and 4 [Table~\ref{FSStable_2D}]. The fitting result with $n = 2, m=3$ gives  \begin{revision}$\nu = 1.980\pm 0.002$\end{revision}. The data points near the critical point for different $W$ and $L$ collapse into a single-parameter scaling function $F(\phi_t)$ for small $\phi_t = u_t L^{1/\nu}$ [an inset of Fig.~\ref{FSS_figure}(c)].

\begingroup
\squeezetable
\begin{table}[htb]
    \caption{\label{FSStable_2D}Finite-size-scaling analyses of the 2D models with and without topology. $L = 120-200$ for model without topology, $L = 160-240$ for model with topology.}
    \begin{ruledtabular}
    \begin{tabular}{ccccc}
        \multicolumn{5}{l}{(a) 2D without topology}\\[0.1cm]
        $n$&$m$&$W_c$&$\nu$&$\Gamma_c$\\
        2&2&1.9629[1.9606,1.9653]&1.933[1.929,1.938]&0.331[0.329,0.333]\\
        2&3&1.9939[1.9917,1.9962]&1.980[1.975,1.985]&0.402[0.400,0.404]\\
        2&4&1.9909[1.9885,1.9934]&1.978[1.974,1.983]&0.399[0.398,0.401]\\
        3&2&2.0314[2.0280,2.0347]&1.676[1.666,1.686]&0.378[0.373,0.382]\\
        4&2&2.0646[2.0614,2.0679]&1.515[1.504,1.525]&0.450[0.447,0.454]\\
        \\[0.1cm]
        \multicolumn{5}{l}{(b) 2D with topology (non-topological direction)}\\[0.1cm]
        $n$&$m$&$W_c$&$\nu$&$\Gamma_c$\\
        2&2&5.9370[5.9359,5.9382]&1.391[1.382,1.400]&9.024[8.980,9.069]\\
        2&3&5.9365[5.9354,5.9377]&1.369[1.357,1.381]&8.977[8.930,9.025]\\
        2&4&5.9348[5.9334,5.9361]&1.317[1.280,1.343]&8.897[8.836,8.955]\\
        3&2&5.9290[5.9273,5.9307]&1.510[1.488,1.533]&8.645[8.577,8.713]\\
        4&2&5.9366[5.9348,5.9383]&1.407[1.386,1.431]&8.916[8.832,8.992]\\
    \end{tabular}
    \end{ruledtabular}
\end{table}
\endgroup

     The 2D disorder-driven phase transitions are studied for zero-energy eigenstates of $h_{\mathrm{T}}^{\mathrm{2D}}$ [Eq.~\eqref{hT_2D}] with $\Delta = 0.1, t_{||} = 0.5, t_{||}' = 0.8, t_{\perp} = 1$. The LEs of $h_{\rm T}$ are calculated along the topological $y$ and non-topological $x$ directions by the transfer matrix method. In the presence of the disorder, the LEs comprise two separate bands [see Fig.~\ref{LEs_3DL16_2DL120}(c) and (d)]. For the LEs along $y$, the lower LE band covers zero in a small $W$ region $(W<W_c^y)$, where $\xi_y$ is divergent and $\nu_y$ takes a non-quantized value. The boundary $W_c^y$ is determined by the linear regression analysis of the LE $\gamma^{(l)}_{\rm max}(W,L)$ on the lower LE band edge, from which we obtain \begin{revision}$W_c^y = 6.1243\pm 0.0006$\end{revision} and $\nu=1$. As shown in Fig.~\ref{FSS_figure}(d), the normalized localization length $\Gamma^{-1}_x(W,L) = \xi_{x}(W,L)/L$ along the non-topological $x$ direction exhibits another delocalization-localization transition within the metallic region $W<W_c^y$ with divergent $\xi_y$ and non-quantized $\nu_y$. The critical point $W_c^x$ and the critical exponent $\nu$ of the transition for $\Gamma_x$ are estimated by the polynomial fitting of the single-parameter scaling function [Eq.~\eqref{scaling_2D}] with $w=(W-W_c^x)/W_c^x$. Since $\Gamma_x(W,L)$ is scale-invariant on the metal-phase side $(W<W_c^x)$, only the data points of $\Gamma_x(W,L)$ for the localized-phase side ($W>W_c^x)$ are fitted by the scaling function. The $\chi^2$ minimization gives \begin{revision}$W_c^x=5.9370 \pm 0.0006$\end{revision} and \begin{revision}$\nu=1.391 \pm 0.005$\end{revision} for $n=m=2$ [see also Table~\ref{FSStable_2D}]. The clear distinction between $W_c^x$ and $W_c^y$ suggests an intermediate 2D QL regime between the critical metal and Anderson localized phases. The estimated critical exponent \begin{revision}$\nu\simeq 1.4$\end{revision} of the 2D metal-to-QL-like transition differs from the 2D Anderson transition $\nu\simeq 2.0$ without 1D topology, indicating distinct critical behavior.

\section{field theory analyses of chiral Anderson transitions}

     The simulation studies of the 3D and 2D chiral symplectic models reveal that, as in the other two chiral symmetry classes~\cite{xiao2023,zhao2024}, the 1D weak topology induces an intermediate quasi-localized (QL) phase or regime between the metallic and Anderson-localized phases. A comparison of universal critical exponents across the three chiral symmetry classes further suggests that the presence, absence, and symplectic nature of time-reversal symmetry influence the quantum criticality of the metal-to-QL transition. Motivated by this observation, we carry out renormalization group (RG) analyses of 2D nonlinear sigma models for all three chiral symmetry classes with 1D weak topology.

     Importantly, our analysis clarifies that the 1D weak topology endows the 2D Anderson transition criticality with spatially anisotropic scaling, while --- contrary to the earlier work of Zhao et al.~\cite{zhao2024} --- it does not support the emergence of a 2D quasi-localized phase adjacent to the metal. The discrepancy arises from the one-loop renormalization of the weak topological term, which was absent in the previous study. In Ref.~\cite{zhao2024}, a strong-coupling fixed point was characterized by a finite conductivity $\sigma_y$ along the topological ($y$) direction and vanishing conductivity $\sigma_x$ along the non-topological ($x$) direction, and the corresponding strong coupling phase was identified  as the 2D quasi-localized phase. When the one-loop renormalization of the weak topological term is incorporated in the analysis, however, this fixed point becomes characterized by vanishing conductivities along {\it both} spatial directions: the strong-coupling phase is a conventional localized phase. Nonetheless, the RG analysis still yields a clear hallmark of the 1D weak topology in the Anderson transition criticality: at the metal-insulator transition, the criticality becomes spatially anisotropic, with a divergent conductivity ratio $\sigma_y/\sigma_x$. This implies that, unlike conventional Anderson transitions, the critical point in the presence of 1D weak topology must be described by spatially anisotropic scaling -- for instance, with distinct scaling dimensions for the length $r_y$ along the topological direction and $r_x$ along the other~\cite{luo2018a,luo2018b}.    
     
        In the following sections, we transform the chiral NLSMs to sine-Gordon (sG) models for all three chiral symmetry classes and derive the corresponding one-loop renormalization group (RG) equations. For the chiral unitary and chiral symplectic classes, the RG flow reveals a weak-coupling fixed point (metal), a strong-coupling fixed point (Anderson insulator), and a saddle point that controls the 2D metal-insulator transition. The chiral orthogonal class, however, does not exhibit a metal-insulator transition within our analysis. Details of the (duality) transformation, and the derivation of the RG equations are given in Appendices B and C, respectively.  

\subsection{2D chiral nonlinear sigma models and their dual sine-Gordon models}
        The Anderson transitions in the chiral symmetry classes can be studied by symmetric-space nonlinear sigma models (NLSM) for a $vN$ by $vN$ unitary matrix $Q$ in the replica limit $[N\rightarrow 0$; in the replica formulation, $N$ is a number of replicated fermions that couple with same quenched disorder]~\cite{gade1991,gade1993,fukui1999,fabrizio2000,konig2012,altland2014,zhao2024,zhao2025},   
\begin{widetext}
\begin{align}
Z &= \int {\cal D} Q \exp [-S[Q]], \label{ft_eq0}  \\
S[Q] & = - \int \frac{d^{2}{\bm r}}{8\pi} 
\bigg\{ \sum_{\mu=r_x,r_y} \Big( 
\sigma_{\mu} {\rm Tr}\big[(Q^{-1}\partial_{\mu}Q)^2]  
+ c_{\mu} {\rm Tr}^2\big[Q^{-1}\partial_{\mu}Q\big] \Big) 
- \chi_y {\rm Tr}\big[Q^{-1}\partial_{r_y} Q\big]\bigg\}. \label{ft_eq1}
\end{align}
\end{widetext}
Here, $\sigma_{\mu}$ and $c_{\mu}$ stand for longitudinal conductivity and Gade constant along $r_{\mu}$ $(\mu=x,y)$, respectively, $\chi_{y}$ is the one-dimensional (1D) weak topological term along the topological direction $r_y$.  The $Q$ field is given by the $N$ by $N$ unitary matrix satisfying $Q^{\dagger}Q=1$ for the chiral unitary (AIII) with $v=1$. For the chiral symplectic (CII) class with $v=1$ and the chiral orthogonal (BDI) class with $v=2$, the $Q$ field is given by an $N$ by $N$ and $2N$ by $2N$ unitary matrix, respectively, subject to additional time-reversal symmetry constraints, 
\begin{align}
&Q^{T} = Q, \ \ \ ({\rm chiral} \!\ \!\ {\rm symplectic}: {\rm CII}),  \nonumber \\
&\tau_y Q^T \tau_y = Q, \ \ \ ({\rm chiral} \!\ \!\ {\rm orthogonal}: {\rm BDI}), \nonumber 
\end{align}
where $\tau_y$ is the 2 by 2 Pauli matrix satisfying $\tau^*_y\tau_y=-1$. Such $Q$ in the chiral unitary,  symplectic and orthogonal classes can be diagonalized as $Q= {\cal M}^{-1} D {\cal M}$, where $D$  is a diagonal matrix with unimodular complex numbers, and ${\cal M}$ is unitary $(U)$ for AIII , orthogonal $(O)$ for CII, and quaternionic self-dual matrix $(V$ satisfying $\tau_y V^* \tau_y=V)$ for BDI: 
\begin{align}
& Q = U^{\dagger} \!\ D \!\ U, \ \ \ ({\rm chiral} \!\ \!\ {\rm unitary}:  {\rm AIII}), \label{notime-a3} \\
& Q = O^{T}\!\ D \!\ O, \ \ \ ({\rm chiral} \!\ \!\ {\rm symplectic}: {\rm CII}), \label{time-c2} \\ 
& Q = V^{\dagger}\!\ D \!\ V, \ \ \ ({\rm chiral} \!\ \!\ {\rm orthogonal}: {\rm BDI}). \label{time-bd1}
\end{align}
As the eigenvalues of the $Q$ field are unimodular, the perturbative $\beta$ function for the conductivity of the chiral non-linear sigma model receives no localization effects for any spatial dimension~\cite{gade1991,gade1993,evers2008}. Instead, the metal-insulator transition is entirely driven by the spatial proliferation of vortex excitations of these eigenvalues~\cite{konig2012,zhao2025}. This mechanism draws parallels with the two-dimensional (2D) Berezinskii-Kosterlitz-Thouless (BKT) transition, and the three-dimensional (3D) superfluid transition driven by vortex-loop proliferation. 
        
       In $d$-dimensional space, vortex excitations generally manifest as closed $(d-2)$-dimensional hypersurfaces. The one-dimensional (1D) weak band topology along the topological ($y$) direction imparts a complex phase factor to each vortex excitation~\cite{tanaka2015,zhao2024,zhao2025}. This phase factor is proportional to the $(d-1)$-dimensional volume of the region enclosed by the vortex hypersurface, projected onto the hyperplane perpendicular to the topological ($y$) direction. Zhao et al. proposed that this phase factor interferes with the proliferation of vortex hypersurfaces that carry a finite projected volume, thereby driving the emergence of a quasi-localized (QL) phase adjacent to the metal~\cite{zhao2024,zhao2025}. They explored this scenario in the two-dimensional (2D) chiral unitary class via a renormalization group (RG) analysis of a generalized U($N$) sine-Gordon (sG) model~\cite{zhao2024}. In three dimensions (3D), using a duality mapping between the U($N$) NLSM and U($N$) type-II superconductors, they drew an analogy between the QL phase induced by the 1D weak topology and the magnetic-flux-line lattice (MFLL) phase in the type-II superconductor under an external magnetic field~\cite{zhao2025}. In the following, we revisit their 2D analysis with the necessary correction and extend it from the chiral unitary class to the other chiral symmetry classes.

     The vortex-driven phase transition in the NLSM for the three chiral symmetry classes, Eq.~(\ref{ft_eq1}), can be studied by using a generalized sine-Gordon model, which is obtained from the sigma model by a duality mapping\cite{konig2012,zhao2024}. In this mapping, a dual matrix field $\Theta({\bm r})$ is introduced as an auxiliary field to describe the vortex degree of freedom. The dual field takes the form of a $vN$ by $vN$ Hermitian matrix. For the chiral symplectic, and chiral orthogonal classes, it is also symmetric under the respective time-reversal symmetry, i.e. Eqs.~(\ref{time-c2},\ref{time-bd1}). The dual matrix field are expanded in terms of  u$(vN)$ Lie algebra generators $T_a$ $(a=0,1,\cdots,D_s-1)$ with real scalar fields $\theta_{a}$ as expansion coefficients,
\begin{align}
\Theta({\bm r}) = \sum^{D_s-1}_{a=0} \theta_{a}({\bm r}) \!\ T_a,  
\end{align}
where $T_0$ is the $vN$ times $vN$ unit matrix, and $T_a$ $(a=1,\cdots)$ are su($vN$) Lie algebra generators.  Here, $D_s$ denotes the number of the u($vN$) Lie algebra generators that are retained under the relevant symmetry constraints, while we call as $D$ the total number of the u($vN$) Lie algebra generators, $D=v^2N^2$. The parameters $(v,D,D_s)$ for each class are $(v,D,D_s)=(1,N^2,N^2)$ for AIII; $(v,D,D_s)=(1,N^2,N(N+1)/2)$ for CII;  $(v,D,D_s)=(2,4N^2,2N^2-N)$ for BDI.  

    With these definitions in place, the generalized sine-Gordon model takes the following explicit form:
\begin{widetext}
\begin{align}
Z = &\int {\cal D}\Theta \exp\Bigg[-2\pi \int d^2{\bm r} \bigg(\frac{\big(\partial_{r_x} \theta_0 - \frac{\chi_y \sqrt{vN}}{8\pi}\big)^2}{\sigma_y+vNc_y} + \frac{\big(\partial_{r_y} \theta_0 \big)^2}{\sigma_x+vNc_x}  \bigg) \nonumber \\
&-2\pi \int d^2{\bm r} \sum^{D_s-1}_{a,b=1} \big[\mathbb{G}(\{\theta_a\})\big]_{a,b}\Big(\frac{(\partial_{r_y} \theta_a)(\partial_{r_y} \theta_b)}{\sigma_x} + \frac{(\partial_{r_x} \theta_a)(\partial_{r_x} \theta_b)}{\sigma_y} -i \frac{(\partial_{r_y} \theta_a)(\partial_{r_x} \theta_b)}{\sqrt{\sigma_x\sigma_y}} + i \frac{(\partial_{r_x} \theta_a)(\partial_{r_y} \theta_b)}{\sqrt{\sigma_x\sigma_y}} \Big) \nonumber \\
&-\frac{\Lambda^2}{2} \int d^2 {\bm r} \!\ {\rm Tr} \ln \Big[\sigma_x\sigma_y \mathbb{I} + \mathbb{P}^2(\{\theta_a\})\Big] + \sum_{\eta\in Z} \!\ y_{\eta} \Lambda^2 \int d^2{\bm r} \int d|p({\bm r})\rangle \!\ e^{i 2v\pi  \!\ \eta  \!\ \sum^{D_s-1}_{a=0} \theta_a({\bm r}) \langle p({\bm r})|T_a|p({\bm r})\rangle}\Bigg]. \label{ft_eq4}
\end{align}
\end{widetext}
In the action, the U(1) component $\theta_0({\bm r})$ of the dual fields plays the role of a 2D electrostatic potential that mediates the logarithmic 2D Coulomb interaction among vortices: the free theory of $\theta_0({\bm r})$ takes the form of the electric part of the 2D Maxwell action in Eq.~(\ref{ft_eq4}). In contrast, the free theory part of the su($vN$) components $\theta_{a}({\bm r})$ $(a=1,\cdots,D_s-1)$ is generally massive, and the sine Gordon model includes interaction terms among these $\mathrm{su}(vN)$ dual fields.

    More specifically, a metric tensor $\mathbb{G}(\{\theta_a\})$ appearing in Eq.~(\ref{ft_eq4}) is a $(D-1)\times (D-1)$ matrix given by,    
\begin{align}
\mathbb{G}(\{\theta_a\}) = \bigg( \mathbb{I} - \frac{i}{\sqrt{\sigma_x\sigma_y}} \mathbb{P}(\{\theta_a\})\bigg)^{-1}, 
\end{align}
where $\mathbb{I}$ is the $(D-1)\times (D-1)$ unit matrix and  $\mathbb{P}(\{\theta_a\})$ is a $(D-1) \times (D-1)$ Hermitian (imaginary antisymmetric) matrix. The antisymmetric matrix is expressed in terms of the structure factor $f_{abc} \equiv -i {\rm Tr}[T_c[T_a,T_b]]$ of the $\mathrm{su}(vN)$ Lie algebra and depends linearly on the su($vN$) dual fields themselves,    
\begin{align}
[\mathbb{P}(\{\theta\})]_{a,b}=-4\pi i \left\{\begin{array}{ll} 
\sum^{D-1}_{c=1}f_{abc} \theta_c, & {\rm for} \,\ {\rm AIII},\\ 
\sum^{D_s-1}_{c=1}f_{abc} \theta_c, & {\rm for} \,\ {\rm CII},\!\ {\rm BDI}.\\ 
\end{array}\right.
\end{align}
Consequently, when the ${\rm Tr}\ln [\sigma_x\sigma_y \mathbb{I}+\mathbb{P}^2]$ term is expanded in powers of the conductivity, the 1st order term $1/(\sigma_x\sigma_y)$ provides a mass term of the $\mathrm{su}(vN)$ components of the dual $\Theta$ field. Similarly, expanding the metric tensor $\mathbb{G}(\{\theta_a\})$  in powers of the conductivity generates higher-order interaction terms among the $\mathrm{su}(vN)$ dual fields. 

      The last term in Eq.~(\ref{ft_eq4}) is the vortex term of the sine-Gordon model, where real positive coefficient $y_{\eta}$ denotes the fugacity of a vortex with vorticity 2$\pi\eta$  ($\eta\in \mathbb{Z}$). The state $| p({\bm r})\rangle$ is an eigenvector of  $Q({\bm r})$ whose eigenvalue hosts a vortex at ${\bm r}$.  $|p({\bm r})\rangle $ is a complex, real and quaternionic real (projective) vectors for the chiral unitary (AIII), chiral symplectic (CII) and chiral orthogonal (BDI) classes, respectively. For the BDI class with $v=2$, the argument of the vortex term carries a factor of $4\pi \eta$, in contrast to $2\pi \eta$ for the AIII and CII classes. This difference arises because, in the BDI class, the argument receives two identical contributions from a pair of the Kramers doublet [see Appendix B].

     In the following RG analysis, we retain only the zero-th order expansion term both in ${\rm Tr}\ln [\sigma_x\sigma_y \mathbb{I}+\mathbb{P}^2]$ and in the metric tensor $\mathbb{G}(\{\theta_a\})$. Furthermore, we consider only the contribution of vortices with the smallest vorticities $\eta=\pm 1$, setting $y_{+1}=y_{-1} \equiv y$.  By applying the gauge transformation $\theta^{\rm old}_0({\bm r}) \rightarrow \theta^{\rm new}_{0}({\bm r})=\theta^{\rm old}_0({\bm r})-\frac{\chi_y \sqrt{vN} r_x}{8\pi} $, the $\chi_y$ dependence is transferred into the argument of the vortex term, yielding 
\begin{widetext}   
\begin{align}
Z = &\int {\cal D}\Theta \exp^{-S_0 - S_y} \nonumber \\
S_0 = & 2\pi \int d^2{\bm r} \bigg(\frac{\big(\partial_{r_x} \theta_0\big)^2}{\sigma_y+vNc_y} + \frac{\big(\partial_{r_y} \theta_0 \big)^2}{\sigma_x+vNc_x}  + \sum^{D_s-1}_{a=1} \Big(\frac{(\partial_{r_x} \theta_a)^2}{\sigma_y} + \frac{(\partial_{r_y} \theta_a)^2}{\sigma_x} \Big)\bigg)  \nonumber \\
S_y = & 2y\Lambda^2 \int d^2{\bm r} \int d|p({\bm r})\rangle \!\ \cos \bigg(\frac{2\pi v}{\sqrt{vN}} \theta_0({\bm r}) + \frac{v \chi_y r_x}{4} + 2\pi v  \!\ \sum^{D_s-1}_{a=1} \theta_a({\bm r}) \langle p({\bm r})|T_a|p({\bm r})\rangle\bigg) \Bigg]. \label{ft_eq5}
\end{align}
\end{widetext}

\subsection{renormalization group equations for the 2D Anderson transition in chiral symmetry classes}
     The sine-Gordon model can be studied using a renormalization group (RG) method. In this approach, each Euclidean dual field $\theta_{a}({\bm r})$ is decomposed into a rapidly varying component and a slowly varying component . A perturbative treatment of the vortex term allows the rapid modes to be integrated out directly, leading to the renormalization of the coupling constant~\cite{giamarchi2004}. Combined with a spatially isotropic length rescaling at each step, this recursive integration of the rapid modes yields the RG equations for the conductivity $\sigma_{\mu}$, the Gade constant $c_{\mu}$, the fugacity parameter $y$, and the weak topological term $\chi_{y}$. In the replica limit, the resulting RG equations for the three chiral symmetry classes are as follows.
\begin{align}
\left\{\begin{array}{l}
\frac{d\sigma_x}{dl} = -\sigma^2_x \Lambda^4 y^2 \!\ {\cal B}_y, \,\ \frac{d\sigma_y}{dl} = -\sigma^2_y \Lambda^4 y^2 \!\ {\cal B}_x, \\
\frac{dc_x}{dl} = -\sigma_x (\sigma_x+2c_x) \Lambda^4 y^2 \!\ {\cal B}_y, \\
 \frac{dc_y}{dl} = -\sigma_y(\sigma_y+2c_y) \Lambda^4 y^2 \!\ {\cal B}_x,  \\
\frac{d\chi_y}{dl} = \chi_y - 4\sigma_y \Lambda^4 y^2 \!\ {\cal C}_x, \,\  \\
 \frac{dy}{dl} = \big(2-\frac{1}{8}\frac{c_x\sigma_y+c_y\sigma_x+2\sigma_x\sigma_y}{\sqrt{\sigma_x\sigma_y}}\big)\!\ y.
 \end{array}\right. \label{a3}
\end{align}
For the chiral symplectic class, they are  
\begin{align}
\left\{\begin{array}{l}
\frac{d\sigma_x}{dl} = -\frac{\pi}{2}\sigma^2_x \Lambda^4 y^2 \!\ {\cal B}_y, \,\ \frac{d\sigma_y}{dl} = -\frac{\pi}{2}\sigma^2_y \Lambda^4 y^2 \!\ {\cal B}_x, \\
\frac{dc_x}{dl} = -\frac{\pi}{2}\sigma_x (\frac{1}{2}\sigma_x+2c_x) \Lambda^4 y^2 \!\ {\cal B}_y, \\
 \frac{dc_y}{dl} = -\frac{\pi}{2}\sigma_y(\frac{1}{2}\sigma_y+2c_y) \Lambda^4 y^2 \!\ {\cal B}_x,  \\
\frac{d\chi_y}{dl} = \chi_y - 2\pi\sigma_y \Lambda^4 y^2 \!\ {\cal C}_x, \,\  \\
 \frac{dy}{dl} = \big(2-\frac{1}{8}\frac{c_x\sigma_y+c_y\sigma_x+2\sigma_x\sigma_y}{\sqrt{\sigma_x\sigma_y}}\big)\!\ y.
 \end{array}\right. \label{c2}
\end{align}
For the chiral orthogonal class, they are  
\begin{align}
\left\{\begin{array}{l}
\frac{d\sigma_x}{dl} = -\frac{4}{\pi}\sigma^2_x \Lambda^4 y^2 \!\ {\cal B}_y, \,\ \frac{d\sigma_y}{dl} = -\frac{4}{\pi}\sigma^2_y \Lambda^4 y^2 \!\ {\cal B}_x, \\
\frac{dc_x}{dl} = -\frac{4}{\pi}\sigma_x (\sigma_x+2c_x) \Lambda^4 y^2 \!\ {\cal B}_y, \\
 \frac{dc_y}{dl} = -\frac{4}{\pi}\sigma_y(\sigma_y+2c_y) \Lambda^4 y^2 \!\ {\cal B}_x,  \\
\frac{d\chi_y}{dl} = \chi_y - \frac{8}{\pi}\sigma_y \Lambda^4 y^2 \!\ {\cal C}_x, \,\  \\
 \frac{dy}{dl} = \big(2-\frac{1}{2}\frac{c_x\sigma_y+c_y\sigma_x+\sigma_x\sigma_y}{\sqrt{\sigma_x\sigma_y}}\big)\!\ y. 
 \end{array}\right. \label{bd1}
\end{align}
Here, ${\cal B}_{\mu}$ and ${\cal C}_{x}$ for the AIII and CII classes $(v=1)$ and the BDI class $(v=2)$ are given by, 
\begin{align}
\left\{\begin{array}{l}
 {\cal B}_\mu =  \frac{v}{2} \Big[I_{0,\mu}+v\frac{dI_{N,\mu}}{dN}\Big|_{N=0}\Big], \,\ \,\ {\rm for}\,\ \,\ \mu=x,y,  \\
 {\cal C}_x = \frac{v}{2}\Big[L_{0,x} + v\frac{dL_{N,x}}{dN}\Big|_{N=0}\Big], \\
 \end{array}\right.
\end{align}
where $I_{N,\mu}$ and $L_{N,x}$ are functions of the coupling constants,  
\begin{align}
\left\{\begin{array}{l}
I_{N,x} = \frac{16\pi}{\epsilon^2_{N,\Lambda}(\sigma_y+Nc_y)}\frac{1-5\tilde{\chi}^2_{N}}{(1+\tilde{\chi}^2_{N})^4},  \\
I_{N,y} = \frac{16\pi}{\epsilon^2_{N,\Lambda}(\sigma_x+Nc_x)}\frac{1}{(1+\tilde{\chi}^2_{N})^3},  \\
L_{N,x} = \frac{16\pi}{\epsilon^{\frac{3}{2}}_{N,\Lambda}(\sigma_y+Nc_y)^{\frac{1}{2}}}\frac{\tilde{\chi}_{N}}{(1+\tilde{\chi}^2_{N})^3},  \\ 
\tilde{\chi}_{N} = \frac{\chi_y}{4\sqrt{\epsilon_{N,\Lambda}(\sigma_y+Nc_y)}},  \\
\epsilon_{N,\Lambda} =  \frac{\Lambda^2}{\sigma_{y}+Nc_y} + \frac{\Lambda^2}{\sigma_{x}+Nc_x}. 
\end{array}\right.
\end{align}
It is worth noting that the 1D weak topological term $\chi_y$ receives a one-loop renormalization proportional to $y^2 {\cal C}_x$. This contribution was not included in our previous work~\cite{zhao2024}, where a strong-coupling fixed point was found to exhibit quasi-localized behavior, characterized by $\sigma_x=0$ and $\sigma_y \ne 0$, and the corresponding strong-coupling phase was identified as a quasi-localized phase. As we show below, however, once this one-loop renormalization is taken into account, the strong-coupling fixed point instead displays conventional localized behavior with $\sigma_x=\sigma_y=0$, implying that the strong-coupling phase should be reinterpreted as a conventional localized phase.  

\subsection{chiral unitary class} 
To analyze the RG phase diagram of Eq.~(\ref{a3}), note first that the following quantities are invariant under Eq.~(\ref{a3}),  
\begin{align}
\lambda_x = \frac{\sigma^2_x}{\sigma_x+c_x}, \,\ \lambda_y = \frac{\sigma^2_y}{\sigma_y+c_y}, \,\ \lambda = \frac{\lambda_x}{\lambda_y}.  
\end{align}
As a result, the RG equation can be reduced to a set of differential equations involving only four independent coupling constants. We choose the followings as the four parameters:   
\begin{align}
&\zeta \equiv \frac{\sigma_x}{\sigma_y}, \,\ K \equiv \frac{1}{8}\frac{c_x\sigma_y+c_y\sigma_x+2\sigma_x\sigma_y}{\sqrt{\sigma_x\sigma_y}}, \nonumber \\ 
& \tilde{y} \equiv y \sqrt{\frac{4\pi(\sigma_y+c_y)\sigma^2_x}{\sigma_y}}, \,\ 
\tilde{\chi} \equiv \frac{\chi_y}{4\Lambda\sqrt{1+\zeta^{-1}}}. 
\end{align}
 The RG equations for these four coupling constants are then given by,  
\begin{align}
\left\{\begin{array}{l}
\frac{dK}{dl} = - K \!\ \tilde{y}^2 \!\ \frac{(3\zeta + \lambda) \tilde{\cal B}_y + (3\lambda+\zeta) \tilde{\cal B}_x}{\zeta + \lambda},  \\ 
\frac{d\zeta}{dl}  = - 2\zeta  \!\ \tilde{y}^2 \!\ \big(\tilde{\cal B}_y - \tilde{\cal B}_x\big), \\
\frac{d\tilde{y}}{dl}  = \tilde{y}\!\ \big(2-K-\tilde{y}^2 \big(\tilde{\cal B}_x+2\tilde{\cal B}_y\big)\big),  \\
\frac{d\tilde{\chi}}{dl} = \tilde{\chi} \!\ \big( 1- \tilde{y}^2 \big(2 \tilde{\cal C}_x + \frac{\tilde{\cal B}_y-\tilde{\cal B}_x}{1+\zeta}\big)\big), 
\end{array}\right. \label{a3_1}
\end{align}
with 
\begin{align}
\left\{\begin{array}{l}
\tilde{\cal B}_x = \frac{\zeta-1+\frac{2\zeta}{\lambda} + (13-4\zeta-17\frac{\zeta}{\lambda}) \tilde{\chi}^2 - 5(2+\zeta-\frac{\zeta}{\lambda}) \tilde{\chi}^4}{(1+\zeta)^3(1+\tilde{\chi}^2)^5},  \\
\tilde{\cal B}_y  =  \frac{2\zeta + (3+2\zeta) \tilde{\chi}^2 
+ \frac{\zeta}{\lambda}\big(1-\zeta - (2+\zeta) \tilde{\chi}^2\big)}{(1+\zeta)^3(1+\tilde{\chi}^2)^4}, \\ 
\tilde{\cal C}_x = \frac{(\zeta - 1) + (\zeta + 2)\tilde{\chi}^2 + \frac{\zeta}{\lambda}(2-\tilde{\chi}^2)}{(1+\zeta)^3(1+\tilde{\chi}^2)^4}, 
\end{array}\right. \label{bxbycx}
\end{align}
and $\lambda \equiv \lambda_x/\lambda_y$.

     Eq.~(\ref{a3_1}) exhibits three distinct fixed-point structures. First, there is a stable weak-coupling fixed region at $\tilde{y}=0$, $K>2$,  finite $\zeta$ and $\tilde{\chi}^{-1}=0$, corresponding to a 2D critical metal phase with finite conductivities $\sigma_x$ and $\sigma_y$. Second, a stable strong-coupling fixed point exists at $\zeta=0$,
 \begin{align} 
 \tilde{\chi}_0&=\frac{1}{2}\sqrt{\frac{35+9\sqrt{17}}{19}}= 0.9741..., \nonumber \\ 
 \tilde{y}_0&=\frac{9}{76}\sqrt{\frac{212909+90075\sqrt{17}}{1957}}=2.046..., 
\end{align}
 with $K=0$, describing a 2D localized phase. Third, there is a saddle fixed point at $\zeta=0$,
 \begin{align}
 \tilde{\chi}_1 &= \frac{\sqrt{7+2\sqrt{10}}}{3} =1.217..., \nonumber \\
 \tilde{y}_1 &= \frac{4}{27}\sqrt{\frac{2(1021+536\sqrt{10})}{31}}=1.961..., \nonumber \\
 K_1 & = \frac{109-10\sqrt{10}}{62} = 1.248..., \nonumber 
\end{align} 
which governs the 2D metal-insulator transition criticality. 

   Importantly, the strong coupling fixed point is characterized by vanishing conductivities in both the topological $(y)$ direction and the non-topological $(x)$ direction. This follows from the fact that the $\beta$ functions of the conductivities are finite and negative at this fixed point, 
\begin{align}
&\frac{d\ln \sigma_x}{dl} = -2 \!\ \tilde{y}^2_0 \!\ \big(\tilde{\cal B}_y\big)\big|_{\zeta=0,\tilde{\chi}=\tilde{\chi}_0} =  - \frac{6(16+3\sqrt{17})}{103}<0, \nonumber \\
&\frac{d\ln \sigma_y}{dl} = -2 \!\ 
\tilde{y}^2_0 \!\ \big(\tilde{\cal B}_x\big)\big|_{\zeta=0,\tilde{\chi}=\tilde{\chi}_0} = \frac{4(-55+9\sqrt{17})}{103} <0.  
\end{align}
In contrast, the Anderson transition criticality is characterized by the saddle-fixed point, at which the conductivity ratio vanishes, $\zeta\equiv \sigma_x/\sigma_y=0$. By linearizing the equation around the saddle fixed point, we obtain four scaling dimensions as $-1.20$, $-0.651 \pm 2.82i$, and $1.37$. The second least irrelevant scaling variable for ``$-1.20...$" is $\zeta$, while the two least irrelevant scaling variables for $-0.651\pm 2.82 i$, and the relevant scaling variable for $1.37...$ are composed of $\delta \tilde{y} = \tilde{y}-\tilde{y}_1$, $\delta K = K-K_1$, and $\delta\tilde{\chi}=\tilde{\chi}-\tilde{\chi}_1$.  From the relevant scaling dimension, the critical exponent of the Anderson transition in the presence of the 1D weak topological term is obtained as $\nu = 1/y_t=1/1.37\simeq0.73$.

\subsection{chiral symplectic class}
The RG equations for the chiral symplectic class [Eq.~(\ref{c2})] also possess two invariant quantities, 
\begin{align}
\lambda_x = \frac{\sigma^2_x}{\sigma_x+2c_x}, \,\ 
\lambda_y = \frac{\sigma^2_y}{\sigma_y+2c_y}, \,\ \lambda = \frac{\lambda_x}{\lambda_y}. 
\end{align}
Thus, as in the AIII class, we choose the following four parameters as the primary coupling constants,
\begin{align}
&\zeta \equiv \frac{\sigma_x}{\sigma_y}, \,\ K \equiv \frac{1}{8}\frac{c_x\sigma_y+c_y\sigma_x+\sigma_x\sigma_y}{\sqrt{\sigma_x\sigma_y}}, \nonumber \\
&\tilde{y} \equiv y \sqrt{\frac{4\pi(\sigma_y+2c_y)\sigma^2_x}{\sigma_y}}, \,\ 
\tilde{\chi} \equiv \frac{\chi_y}{4\Lambda\sqrt{1+\zeta^{-1}}}. 
\end{align}
Unlike the AIII class, however, the RG equations for these four parameters depend not only on themselves but also on $\sigma_y$, 
\begin{align}
\left\{\begin{array}{l}
\frac{dK}{dl} = - \frac{\pi}{4} K \!\ \tilde{y}^2 \!\ \frac{(3\zeta + \lambda) \tilde{\cal B}^{\prime}_y  + (3\lambda+\zeta) \tilde{\cal B}^{\prime}_x}{\zeta + \lambda},  \\ 
\frac{d\zeta}{dl}  = - \frac{\pi}{2}\zeta \!\ \tilde{y}^2 \!\ \big(\tilde{\cal B}^{\prime}_y - \tilde{\cal B}^{\prime}_x\big) , \\
\frac{d\tilde{y}}{dl}  = \tilde{y} \!\ \big(2-K -\frac{1}{8} \sigma_y \sqrt{\zeta} -\frac{\pi}{4}\tilde{y}^2 \!\ \big(\tilde{\cal B}^{\prime}_x+2\tilde{\cal B}^{\prime}_y \big)\big), \\
\frac{d\tilde{\chi}}{dl} = \tilde{\chi}\!\ \big(1- \frac{\pi}{4}\tilde{y}^2 \big(2 \tilde{\cal C}^{\prime}_x   + \frac{1}{1+\zeta}\big(\tilde{\cal B}^{\prime}_y-\tilde{\cal B}^{\prime}_x \big) \big)\big),  \\
\end{array}\right. \label{c2_2}
\end{align} 
where $\tilde{\cal B}^{\prime}_{\mu}$ $(\mu=x,y)$ and $\tilde{\cal C}^{\prime}_x$ are given by  
\begin{align}
\left\{\begin{array}{l}
\tilde{\cal B}^{\prime}_{x}=\tilde{\cal B}_x + \frac{\lambda_y}{\sigma_y} \tilde{\cal F}_x, \\
\tilde{\cal B}^{\prime}_{y}=\tilde{\cal B}_y + \frac{\lambda_y}{\sigma_y}  \tilde{\cal F}_y, \\ 
\tilde{\cal C}^{\prime}_x =\tilde{\cal C}_x + \frac{\lambda_y}{\sigma_y}  \tilde{\cal F}_y, \\
\tilde{\cal F}_x = \frac{1}{(1+\zeta)^2}\frac{1-5\tilde{\chi}^2}{(1+\tilde{\chi}^2)^4}, \\ 
\tilde{\cal F}_y = \frac{1}{(1+\zeta)^2}\frac{1}{(1+\tilde{\chi}^2)^3}, \\
\end{array}\right. \label{fxfy}
\end{align}
$\tilde{\cal B}_{\mu}$ and $\tilde{\cal C}_x$ are defined in Eq.~(\ref{bxbycx}).  Consequently, Eq.~(\ref{c2_2}) must be solved together with the RG equation for $\sigma_y$: 
\begin{align}
\frac{d\sigma_y}{dl} = - \frac{\pi}{2} \tilde{y}^2  \sigma_y \tilde{\cal B}^{\prime}_x.  \label{c2_3}
\end{align}

Eqs.~(\ref{c2_2},\ref{c2_3}) admit a weak-coupling fixed region at $2-K-\frac{1}{8}\sigma_y \sqrt{\zeta}<0$, $\tilde{y}=0$, $\tilde{\chi}^{-1}=0$ and finite $\zeta$ and $\sigma_y$, corresponding to a critical metallic phase. They also exhibit a strong coupling fixed point at $\sigma_x=\sigma_y=0$, $K=\tilde{\chi}=0$ with finite $\zeta$ and finite $\tilde{Y}$ where, 
\begin{align}
\tilde{Y}\equiv \frac{\tilde{y}}{\sqrt{\sigma_y}}= y\!\ \zeta\sqrt{4\pi(\sigma_y+2c_y)}.
\end{align}
The RG equation for $\tilde{Y}$ reads, 
\begin{align}
&\frac{d\tilde{Y}}{d\ln b} = \tilde{Y}\big(2-K - \frac{1}{8}\sigma_y \sqrt{\zeta} 
-\frac{\pi}{2}\tilde{Y}^2 \sigma_y \tilde{\cal B}^{\prime}_y\big).
\end{align}
The stability of the strong coupling fixed point can be analyzed by linearizing the RG equations for $(\tilde{Y},\zeta, K, \tilde{\chi}, \sigma_y)$ around the fixed point $(\tilde{Y},\zeta, K, \tilde{\chi}, \sigma_y)=(\tilde{Y}_0,\zeta, 0, 0, 0)$, 
\begin{align}
\left\{\begin{array}{l}
\frac{d\tilde{Y}}{d\ln b}  \simeq \tilde{Y} \Big[2 
- \frac{\pi}{2} \tilde{Y}^2 \frac{\lambda_y}{(1+\zeta)^2}\Big] = \frac{2 \tilde{Y}(\tilde{Y}_0-\tilde{Y})(\tilde{Y}_0+\tilde{Y})}{\tilde{Y}^2_0},  \\
\frac{d\zeta}{d\ln b}  \simeq -\frac{\pi}{2} \zeta \tilde{Y}^2_0 (\tilde{\cal B}_y-\tilde{\cal B}_x) \sigma_y \simeq 0,  \\
\frac{dK}{d\ln b}  \simeq - 4K, \,\ 
\frac{d\tilde{\chi}}{d\ln b} \simeq  - \tilde{\chi}, \,\
\frac{d\sigma_y}{d\ln b} \simeq -2\sigma_y, \\   
\end{array}\right.
\end{align}
with $\tilde{Y}_0 \equiv 2(1+\zeta)/\sqrt{\pi \lambda_y}$.  Since the fixed point is characterized by vanishing conductivities along {\it both} spatial directions, the strong-coupling phase is identified as a localized phase. 

The quantum criticality of the metal-insulator transition is governed by a saddle fixed point at $\sigma^{-1}_y=0$, $\zeta=0$ with $\sigma_y \sqrt{\zeta}=0$, $K =K_1$, $\tilde{y}=\tilde{y}_1$, and $\tilde{\chi}=\tilde{\chi}_1$, where   
\begin{align}
\tilde{\chi}_1 &= \frac{\sqrt{7+2\sqrt{10}}}{3} = 1.217...,\nonumber \\
\tilde{y}_1 &= \frac{8}{27}\sqrt{\frac{2(1021+536\sqrt{10})}{31\pi}} = 2.213..., \nonumber \\ 
K_1 & = \frac{109-10\sqrt{10}}{62} = 1.248... . \nonumber 
\end{align}
The RG equations for $\sigma^{-1}_y$ and $\sigma_y \sqrt{\zeta}$ are 
\begin{align}
\frac{d\sigma^{-1}_y}{d\ln b} &= \frac{\pi}{2} \tilde{y}^2 \sigma^{-1}_y 
\tilde{\cal B}^{\prime}_x, \nonumber \\
\frac{d (\sigma_y \sqrt{\zeta})}{d\ln b} &= -\frac{\pi}{4} \!\ \tilde{y}^2 \sigma_y \sqrt{\zeta} \!\  
\big(\tilde{\cal B}^{\prime}_x+\tilde{\cal B}^{\prime}_y\big). \nonumber 
\end{align}
Around this fixed point, both $\sigma^{-1}_{y}$ and $\sigma_y \sqrt{\zeta}$ flow attractively to zero,  
\begin{align}
&\big(\tilde{\cal B}^{\prime}_x\big)\big|_{\zeta=0,\tilde{\chi}=\tilde{\chi}_1,\sigma^{-1}_{y}=0} = \frac{-247+14\sqrt{10}}{5184} <0, \nonumber \\
&\big(\tilde{\cal B}^{\prime}_x +\tilde{\cal B}^{\prime}_y\big)\big|_{\zeta=0,\tilde{\chi}=\tilde{\chi}_1,\sigma^{-1}_y=0} =  \frac{247-14\sqrt{10}}{2592}>0. \nonumber 
\end{align}
Linearizing the equation for the remaining three variables, $\delta \tilde{y} = \tilde{y}-\tilde{y}_1$, $\delta K = K-K_1$, $\delta\tilde{\chi}=\tilde{\chi}-\tilde{\chi}_1$, around this fixed point yields three scaling dimensions: $-1.00...\pm 2.76i$, and $1.329$. The universal critical exponent for the phase transition is thus $\nu = 1/y_t=1/1.329\simeq 0.752$. Notably, the saddle fixed point is spatially anisotropic with $\zeta=\sigma_x/\sigma_y\to 0$, suggesting that the Anderson transition criticality in this class is characterized by the anisotropic spatial scaling.      

\subsection{chiral orthogonal class}
     The RG equations for the chiral orthogonal class [Eq.~(\ref{bd1})] possess two invariant quantities, 
\begin{align}
\lambda_x = \frac{\sigma^2_x}{\sigma_x+c_x}, \,\ \lambda_y = \frac{\sigma^2_y}{\sigma_y+c_y}, \,\ \lambda = \frac{\lambda_x}{\lambda_y}.  
\end{align}
As in the previous cases, we choose the following four coupling constants as the primary parameters,
\begin{align}
&\zeta \equiv \frac{\sigma_x}{\sigma_y}, \,\ K \equiv \frac{1}{2}\frac{c_x\sigma_y+c_y\sigma_x+2\sigma_x\sigma_y}{\sqrt{\sigma_x\sigma_y}}, \nonumber \\ 
&\tilde{y} \equiv y \sqrt{\frac{4\pi(\sigma_y+c_y)\sigma^2_x}{\sigma_y}}, \,\ 
\tilde{\chi} \equiv \frac{\chi_y}{2\Lambda\sqrt{1+\zeta^{-1}}}. 
\end{align}
Much like in the CII class, the RG equations for these four depend not only on themselves but also on $\sigma_y$, 
\begin{align}
\left\{\begin{array}{l}
\frac{dK}{dl} = - \frac{8}{\pi} K \!\ \tilde{y}^2 \!\ \frac{(3\zeta + \lambda) \tilde{\cal B}^{\prime\prime}_y  + (3\lambda+\zeta) \tilde{\cal B}^{\prime\prime}_x}{\zeta + \lambda},  \\ 
\frac{d\zeta}{dl}  = - \frac{16}{\pi}\zeta \!\ \tilde{y}^2 \!\ \big(\tilde{\cal B}^{\prime\prime}_y - \tilde{\cal B}^{\prime\prime}_x\big) ,  \\
\frac{d\tilde{y}}{dl} = \tilde{y} \!\ \big(2-K +\frac{1}{2} \sigma_y \sqrt{\zeta} -\frac{8}{\pi}\tilde{y}^2 \!\ (\tilde{\cal B}^{\prime\prime}_x+2\tilde{\cal B}^{\prime\prime}_y) \big),  \\
\frac{d\tilde{\chi}}{dl} = \tilde{\chi}\!\ \big( 1- \frac{8}{\pi}\tilde{y}^2 \big(2\tilde{\cal C}^{\prime\prime}_x  + \frac{1}{1+\zeta}(\tilde{\cal B}^{\prime\prime}_y-\tilde{\cal B}^{\prime\prime}_x) \big)\big), \\
\frac{d\sigma_y}{dl} = - \frac{16}{\pi} \tilde{y}^2 \!\ \sigma_y \tilde{\cal  B}^{\prime\prime}_x. \\
\end{array}\right. \label{bd1_2}
\end{align}
where $\tilde{\cal B}^{\prime\prime}_{\mu}$ $(\mu=x,y)$ and $\tilde{\cal C}^{\prime\prime}_x$ are given by   
\begin{align}
\left\{\begin{array}{l}
\tilde{\cal B}^{\prime\prime}_{x}=2\tilde{\cal B}_x - \frac{\lambda_y}{\sigma_y} \tilde{\cal F}_x, \\
\tilde{\cal B}^{\prime\prime}_{y}=2\tilde{\cal B}_y - \frac{\lambda_y}{\sigma_y}  \tilde{\cal F}_y, \\ 
\tilde{\cal C}^{\prime\prime}_x =2\tilde{\cal C}_x - \frac{\lambda_y}{\sigma_y}  \tilde{\cal F}_y, \\
\end{array}\right. 
\end{align}
with  $\tilde{\cal B}_{\mu}$, $\tilde{\cal F}_{\mu}$ and $\tilde{\cal C}_x$ defined in Eqs.~(\ref{bxbycx},\ref{fxfy}).

    According to numerical solutions, Eq.~(\ref{bd1_2}) does not exhibit any phase transition. Instead, the coupling constants always flow into a metallic phase with finite conductivities, finite $K$ and finite conductivity ratio $\zeta$.  To properly describe the 2D Anderson transition in chiral orthogonal class, it may be necessary to include a finite mass term for the su($2N$) components $\theta_a({\bm r})$  $(a=1,2,\cdots,2N^2-N-1)$ in the sine-Gordon action, --- a contribution that is entirely neglected in the zero-th order expansion of the metric tensor $\mathbb{P}(\{\theta_a\})$ with respect to the conductivity. We leave this investigation for future work.

\section{Summary and Discussion}

      In this article, we have studied the nature of the Anderson transition in chiral symmetry classes with one-dimensional (1D) weak topology, using numerical simulations and field-theory analyses. In the simulation work, we investigate disorder-driven quantum phase transitions in the chiral symplectic class (class CII) based on 3D models with and without 1D weak topology. We determine the universal critical exponent for the 3D Anderson transition in this symmetry class without 1D weak topology [Fig.~\ref{pd_3D}(a)] and find that the estimated exponent aligns with previous results for a 3D non-Hermitian class AII model~\cite{luo2022}.

     We also study the effect of a weak 1D band topology on the Anderson transition, based on the finite-size scaling analysis that assumes the spatially-isotropic scaling. In 3D, we confirm that the weak topology splits the delocalization-localization transition into a two-step process, giving rise to an intermediate quasi-localized (QL) phase between the diffusive metal and Anderson insulator phases [Fig.~\ref{pd_3D}(b)]. This finding, together with earlier reports of QL phases in chiral orthogonal and unitary classes~\cite{xiao2023}, is consistent with a Berry-phase mechanism for the emergence of the intermediate QL phase~\cite{zhao2024,zhao2025}.

     We further extend the numerical analysis to 2D systems. In the model without 1D topology, the normalized localization length shows a direct transition from a critical metal phase to the Anderson localized phase, with a critical exponent $\nu\simeq 2.0$. In contrast, when the weak 1D topology is present, the localization lengths along the topological and non-topological directions yield two distinct transitions in the finite-size fitting. Their clear separation suggests an intermediate 2D QL regime between the critical metal and the Anderson localized phase. Moreover, the transition associated with the non-topological direction gives \begin{revision}$\nu\simeq 1.4$\end{revision}, distinct from the value obtained in the model without 1D topology. These numerical results indicate that the weak 1D topology does not merely shift the transition point and exponent, but qualitatively changes the localization process by introducing strong spatial anisotropy.

     The 2D field theory study offers a complementary perspective on the role of the 1D weak topology in the chiral symmetry classes. With the inclusion of the one-loop renormalization of the weak topological term, the strong-coupling fixed point is characterized by vanishing conductivities in both spatial directions, indicating a conventional localized phase rather than a quasi-localized one. Meanwhile, the saddle fixed point governing the metal-insulator transition satisfies $\sigma_x/\sigma_y\rightarrow 0$, signaling a spatially anisotropic scaling feature of the metal-insulator transition point.  That says, instead of supporting the QL strong-coupling fixed point proposed in the earlier work, the RG equations point to anisotropic spatial scaling as the key signature of the 1D weak topology at the Anderson transition criticality. 

    The discrepancy between this 2D RG scenario and the numerically observed QL regime may stem from the scaling assumption adopted in the present FSS analysis. In the transfer matrix study, the normalized localization length $\Lambda = \xi/L$ is defined using the correlation length $\xi$ (the inverse of the smallest Lyapunov exponent) along the quasi-one-dimensional direction and the finite system size $L$ along the transverse direction, or vice versa. This quantity is scale-invariant only at critical points where the two spatial directions share the same scaling dimension. For critical points with spatially anisotropic scaling~\cite{cardy1996}, an appropriate scale-invariant quantity should instead involve a nontrivial power of the transverse system size~\cite{luo2018a,luo2018b}. The isotropic scaling assumption employed here may therefore affect the transition identified along the non-topological direction and, consequently, the extent of the numerically resolved QL regime. In this light, the RG scenario offers a new perspective on how weak topology modifies critical behavior and calls for further investigation of the underlying scaling structure.

\begin{acknowledgments}
    We thank Xunlong Luo, Kohei Kawabata, and Tomi Ohtsuki for discussions. The work was supported by the National Basic Research Programs of China (No. 2024YFA1409000) and the National Natural Science Foundation of China (No. 12074008 and No. 12474150).
\end{acknowledgments}

\appendix
\section{Density of states of 3D non-topological model}

    The density of states (DOS) of the 3D non-topological model $H_{\rm NT}$ is calculated using the kernel polynomial method (KPM)~\cite{Alexander2006KPM}, The DOS provides a complementary information about the model. The lower bound $-E_b$ and upper bound $E_{\mathrm{b}}$ of eigenvalues of $H_{\rm NT}$ are estimated by the Lanczos algorithm~\cite{lanczos1950iteration}. We rescale the Hamiltonian by a factor $E_{\rm b}/(1-\epsilon)$ with $\epsilon=0.01$, so that eigenvalues $\tilde{E}$ of the rescaled Hamiltonian is in a range of $[-1,1]$. The DOS as a function of $\tilde{E}$ is expanded by the Chebyshev polynomials of the first kind $T_n(x) =\cos(n\arccos x)$,
    \begin{equation}
        \tilde{\rho}(\tilde{E}) = \frac{1}{\pi\sqrt{1-\tilde{E}^2}}\left[ g_0\mu_0T_0 + 2\sum_{n = 1}^{N-1} g_n\mu_nT_n(\tilde E) \right].
    \end{equation}
    Here $\mu_n$ is the expansion moment; $g_n$ is the Jackson kernel~\cite{jackson1912approximation}, which suppresses the Gibbs oscillations due to a finite truncation of the order of expansion. The moments are evaluated stochastically as~\cite{Skilling1989MaximumEntropy,Drabold1993MaximumEntropy,Silver1994Stochastic}
    \begin{equation}\label{moments}
        \mu_n=\frac{1}{D}\mathrm{Tr}[T_n(\tilde H)] \approx \frac{1}{DR}\sum_{r=0}^{R-1}\langle r|T_n(\tilde H)|r\rangle,
    \end{equation}
    where $D$ is the dimension of $H$, $|r\rangle$ is a random vector, and $R$ is the number of  random vectors. We use $R = 30$ in all calculations.

    \begin{figure}
        \centering
        \includegraphics[width=\linewidth]{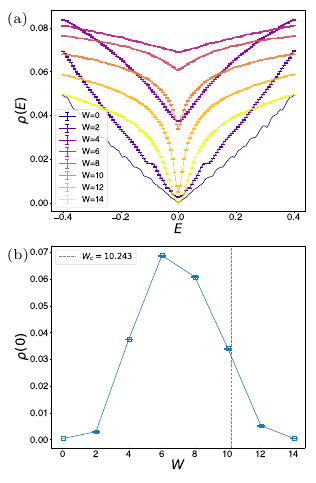}
        \caption{\label{DOS_CII_3D_wo}Density of states (DOS) of the 3D nontopological model $H_{\mathrm{NT}}$ with $\Delta = D = F = t_{\perp} = t_{||} = 1$, $t_{||}' = 0.5$. (a) The DOS as a function of energy in the range $E \in [-0.4,0.4]$. (b) The DOS at zero energy $\rho(E = 0)$ as a function of the disorder strength $W$. The data points with error bars for $W>0$ are averaged over 5 disorder realizations of system size $L^3 = 80^3$ and truncation order $N = 4000$. For $W=0$, the DOS is calculated with a larger system size $L^3 = 200^3$, where a smaller truncation $N = 3000$ is used to suppress artificial oscillations. Grey dashed line in (b) marks the Anderson transition point $W = W_c$ according to the fitting results in Table~\ref{FSStable_3D}.}
    \end{figure}

    Fig.~\ref{DOS_CII_3D_wo} shows the DOS near zero energy for different disorder strengths. In the clean limit, the model exhibits semimetallic low-energy behavior, with a pseudogap around $E=0$: $\rho(0)\simeq 0$. As the disorder strength increases, the pseudogap is gradually filled with states and the system develops a finite zero-energy DOS. With stronger disorder, $\rho(0)$ decreases after reaching a maximum. This results in the nonmonotonic $W$-dependence of $\rho(0)$, as shown in Fig.~\ref{DOS_CII_3D_wo}(b). At the numerical resolution used, $\rho(0)$ remains finite at the Anderson transition point [\begin{revision}$W_c=10.243$\end{revision}, determined by the FSS analysis of the Q1D localization length], indicating that the Anderson transition occurs in a regime with a finite density of states.

\section{duality mapping to generalized sine-Gordon models}

     To establish a duality mapping between the sigma model and the generalized sine-Gordon model, let us first consider a saddle-point solution of the sigma model in which an eigenvalue of $Q({\bm r})$ exhibits a vortex excitation. Such saddle-point solution is generally described by the following ansatz for the classes AIII and CII~\cite{konig2012,zhao2024,zhao2025}, 
\begin{align}
Q({\bm r}) = 1 + \big(e^{i\phi({\bm r})} - 1 \big) |p({\bm r})\rangle \langle p({\bm r})|. 
\end{align}
Here, an eigenvector $|p(\bm r)\rangle$ for an eigenvalue $e^{i\phi({\bm r})}$ is a complex and real projective vector for the classes AIII and CII, respectively. For the class BDI, eigenvectors of $Q({\bm r})$ for $e^{i\phi({\bm r})}$ are doubly degenerate; we thus use the following ansatz of $Q({\bm r})$ for the class BDI, 
\begin{align}
Q({\bm r}) = 1 + \big(e^{i\phi({\bm r})} - 1 \big) \sum_{\sigma=\pm}|p({\bm r}),\sigma\rangle \langle p({\bm r}),\sigma|, 
\end{align}
with $|p,+\rangle \langle p,+| = \tau_y (|p,-\rangle \langle p,-|)^T \tau_y$ and $\langle p,+ |p,-\rangle =0$. For these ans\"atze to realize a saddle point of (the real part of) $S[Q]$ in Eq.~(\ref{ft_eq1}),  they must satisfy $(\sigma_{\mu} + v c_{\mu}) \partial^2_{\mu} \phi({\bm r}) =0$, and $|p({\bm r})\rangle$ must be independent of ${\bm r}$ except for at the vortex core [$v=1$ for AIII and CII, and $v=2$ for BDI]. After normalizing $r_x$ and $r_y$ by $\sqrt{\sigma_x+v c_x}$ and $\sqrt{\sigma_y+vc_y}$  respectively, the Poisson equation gives a vortex solution at ${\bm r}={\bm r}_0$ with the vorticity $\eta \in \mathbb{Z}$, $\phi(r_x,r_y)= \eta \!\ \tan^{-1}[(r_y-r_{y,0})/(r_x-r_{x,0})]$.  To take a Gaussian integral around a saddle-point solution with multiple vortices, we introduce a Euclidean field $h_{\mu}({\bm r})\equiv -i Q^{-1}\partial_{\mu}Q$, and integrate over $h_{\mu}({\bm r})$ under the following condition imposed on $F_{xy}({\bm r})\equiv \partial_{r_x}h_{y}-\partial_{r_y}h_x+i [h_x,h_y]$~\cite{konig2012,zhao2024,zhao2025},
\begin{widetext}
\begin{align}
F_{xy}({\bm r}) = J_n({\bm r}) \equiv \sum^n_{j=1} 2\pi \eta_j \!\ \delta^2({\bm r}-{\bm r}_j)  \left\{ \begin{array}{ll} 
|p({\bm r}_j)\rangle 
\langle p({\bm r}_j)| &  \!\ \!\ {\rm for} \,\ {\rm AIII}, \!\ {\rm CII}  \\ 
\sum_{\sigma=\pm } |p({\bm r}_j),\sigma\rangle 
\langle p({\bm r}_j),\sigma| &  \!\ \!\ {\rm for} \,\  {\rm BDI}.  \\
\end{array} \right.
\end{align}
Here, ${\bm r}_j$ and $\eta_j$  are the 2D coordinate, and vorticity of the $j$-th vortex, while $|p({\bm r}_j)\rangle$ ($|p(\bm r_j),\pm\rangle$ for BDI) is an eigenvector at the vortex core for the $j$-th vortex $(j=1,\cdots,n)$.  Namely, ${\bm r}_j$, $\eta_j$ and $|p({\bm r}_j)\rangle$ as well as $h_{\mu}({\bm r})$ are integral variables of the partition function, 
\begin{align}
Z &= \int {\cal D} h_{\mu} \!\ e^{S[h_{\mu}]} \!\ \sum^{\infty}_{n=0} \frac{1}{n!} 
\prod^n_{j=1} \Big(\int {\cal D}^2{\bm r}_j \int d|p({\bm r}_j)\rangle \sum_{\eta_j\in \mathbb{Z}} y_{\eta_j} \Big) 
\prod_{\bm r} 
\delta(F_{xy}({\bm r}) - J_{n}({\bm r})) , \label{ft_eq2} \\
S[h] & = \int \frac{d^{2}{\bm r}}{8\pi} \bigg\{ \sum_{\mu=r_x,r_y} \Big( 
\sigma_{\mu} {\rm Tr}\big[h^2_{\mu}]  
+ c_{\mu} {\rm Tr}^2\big[h_{\mu}\big] \Big) 
+ i\chi_y \!\ {\rm Tr}\big[h_{r_y}\big]\bigg\}. \label{ft_eq3}
\end{align}
\end{widetext}
Here we introduce a real positive parameter $y_{\eta}$ as the fugacity parameter of the vortex with vorticity $\eta$.  $\sigma_\mu$ , $c_{\mu}$ and $\chi_y$ in $S[h]$ are also normalized from Eq.~(\ref{ft_eq1}) to Eq.~(\ref{ft_eq3}) : $\sigma^{\rm new}_x = \sigma_x \sqrt{\frac{\sigma_y+vc_y}{\sigma_x+vc_x}}$, $\sigma^{\rm new}_y = \sigma_y \sqrt{\frac{\sigma_x+vc_x}{\sigma_y+vc_y}}$, $\chi^{\rm new}_y=\sqrt{\sigma_x+vc_x}\!\ \chi_y$. ${\cal D}^2{\bm r} \equiv \Lambda^{2} d^2{\bm r}$ is dimensionless with a ultraviolet cutoff $\Lambda^{-1}$ in the real space.  An integral over an eigenvector $|p\rangle$ of $Q$ at each vortex core is taken over the $\mathbb{CP}^{N-1}$ space for AIII, the $\mathbb{RP}^{N-1}$ space for CII, and  the $\mathbb{HP}^{N-1}$ space for BDI, 
 \begin{widetext}
\begin{align}
\int d|p\rangle = \left\{\begin{array}{ll}
\frac{1}{\pi} \int^{+\infty}_{-\infty} \prod^N_{i=1} d{\rm Re}p_i d{\rm Im}p_i 
\!\ \delta \big(\sum^N_{i=1}|p_i|^2-1\big) & {\rm for}\,\ {\rm AIII}, \\ 
\int^{+\infty}_{-\infty} \prod^N_{i=1} dx_i\!\ \delta\big(\sum^N_{i=1} x^2_i -1 \big) & {\rm for} \,\ 
{\rm CII},  \\
\frac{1}{\pi^2} \int^{+\infty}_{-\infty} \prod^N_{i=1}\prod_{\tau=\pm} 
d{\rm Re}p_{i,\tau} d{\rm Im}p_{i,\tau}\!\ \delta \big(\sum^N_{i=1} \sum_{\tau=\pm}|p_{i,\sigma}|^2-1\big) & {\rm for}\,\ {\rm BDI}. \\ 
\end{array}\right. 
\end{align}  
\end{widetext}

   The local constraint can be treated with an auxiliary matrix field $\Theta({\bm r})$, 
\begin{align}
&\prod_{\bm r}\delta(F_{xy}({\bm r}) - J_{n}({\bm r})) = \nonumber \\
&\!\ \int {\cal D}{\Theta}({\bm r}) 
\exp \Big[-i\int d^2{\bm r}{\rm Tr}\big[\Theta({\bm r})(F_{xy}({\bm r})-J_n({\bm r}))\big]\Big]. 
\end{align}
As $F_{xy}({\bm r})$ and $J_{n}({\bm r})$ are Hermitian, real symmetric, and quaternionic self-dual matrices for the classes AIII, CII, and BDI, respectively, so is the auxiliary field $\Theta({\bm r})$; 
\begin{align} 
\left\{\begin{array}{ll} 
\Theta^{\dagger}=\Theta,   &  {\rm for} \,\ {\rm AIII}, \\ 
\Theta^{\dagger}=\Theta, \!\ \Theta^* = \Theta,  &  {\rm for} \,\ {\rm CII}, \\ 
\Theta^{\dagger}=\Theta,  \!\ \tau_y \Theta^{*} \tau_y = \Theta, &  {\rm for} \,\ {\rm BDI}. \\ 
\end{array}\right.
\end{align}
Here, $\Theta$ for the classes AIII and CII are $N$ by $N$ matrices, while they are $2N$ by $2N$ matrices for the class BDI. The auxiliary fields thus introduced can be expanded in terms of $\mathrm{u}(N)$ linear algebra $T_a$ $(a=0,2,\cdots,N^2-1)$ for the class AIII, real symmetric components of $\mathrm{u}(N)$ for the class CII, quaternionic self-dual components of $\mathrm{u}(2N)$ for the class BDI, 
\begin{align}
\Theta({\bm r})=\left\{\begin{array}{ll} 
\!\ \sum^{N^2-1}_{a=0} \theta_{a}({\bm r}) \!\ T_a,   &  {\rm for} \,\ {\rm AIII}, \\ 
\!\ \sum^{D_s-1}_{a=0} \theta_a({\bm r}) \!\ T^s_a & {\rm for} \,\ {\rm  CII}, \\ 
\!\ \sum^{D_s-1}_{a=0} \theta_a({\bm r}) \!\ T^s_a, &  {\rm for} \,\ {\rm BDI}. \\ 
\end{array}\right.
\end{align}
Thereby, $T^s_a$ for the class CII stands for a real symmetric component of $\mathrm{u}(N)$ lie algebra $[(T^s_a)^T=T^s_a]$ with $D_s\equiv \frac{N(N+1)}{2}$, whereas $T^s_{a}$ for the class BDI is a quaternionic self-dual component of $\mathrm{u}(2N)$ lie algebra  $[\tau_y (T^s_a)^T \tau_y=T^s_a]$ with $D_s \equiv 2N^2-N$.  For all the three chiral classes, $T_{0}$ as well as $T^s_{0}$ are chosen to be the unit matrix, $T_{0}=T^s_0 \equiv \frac{1}{\sqrt{vN}}\mathbb{I}_{vN\times vN}$ with $v=1$ for AIII and CII, and $v=2$ for BDI. The lie algebra is  normalized with ${\rm Tr}[T_a T_b]={\rm Tr}[T^s_a T^s_b]=\delta_{a,b}$. Note that u$(N)$ lie algebra thus normalized has the following Fierz identity, $\sum^{D}_{a=0} (T_a)_{ij} (T_a)_{lm} = \delta_{im}\delta_{jl}$. Likewise, its real symmetric components and quaternionic self-dual components of u$(2N)$ lie algebra also have similar identities. 
\begin{align}
&\sum^{D_s}_{a=0} (T^s_a)_{ij} (T^s_a)_{lm}  = \nonumber \\
& \left\{\begin{array}{ll} 
 \frac{1}{2}\big(\delta_{im}\delta_{jl} + 
 \delta_{il}\delta_{jm}\big) & {\rm for} \,\ {\rm CII}, \\
\frac{1}{2}\big(\delta_{im}\delta_{jl} + (\tau_y)_{il}(\tau_y)_{jm}\big)  & {\rm for} \,\ {\rm BDI}. \\
\end{array}\right. \label{fierz_2}
\end{align}

      Meanwhile, for all the chiral symmetry classes, $h_{\mu}({\bm r}) \equiv -i Q^{-1}({\bm r})\partial_{\mu}Q({\bm r})$ has both symmetric and antisymmetric components of the u$(vN)$ lie algebra under the transposition. Thus, it is expanded entirely in terms of the $\mathrm{u}(vN)$ lie algebra,
\begin{align}
h_{\mu}({\bm r}) = \sum^{D-1}_{a=0} \pi_{\mu,a}({\bm r}) T_{a},
\end{align}
with $D=N^2$ for the classes AIII and CII, and $D=4N^2$ for the class BDI. The Gaussian integration over $\pi_{\mu,a}({\bm r})$ $(\mu=r_x, r_y, a=0,1,\cdots,D-1)$  completes a duality mapping to the sine-Gordon action, Eq.~(\ref{ft_eq4}).

\section{derivation of RG equations for three chiral symmetry classes}

    The sine-Gordon models for the chiral symmetry classes, Eq.~(\ref{ft_eq5}), can be studied using an authentic renormalization group (RG) method. Thereby, the dual field $\theta_{a}$ $(a=0,1,\cdots)$ is decomposed into a fast mode $\theta^{>}_a$ and a slow mode $\theta^{<}_a$, and so is the free part of the action $S_0$, $S_0=S^<_0+S^>_0$.
\begin{align}
S^{<}_{0} &= \int_{\epsilon_{vN,{\bm k}}<\epsilon_{vN,\Lambda}e^{-2dl} } 
\frac{d^2{\bm k}}{2\pi}  \epsilon_{vN,{\bm k}} \,\ \theta^{<}_0({\bm k}) \theta^{<}_0(-{\bm k}) \nonumber \\
& + \int_{\epsilon_{0,{\bm k}}<\epsilon_{0,\Lambda}e^{-2dl} } 
\frac{d^2{\bm k}}{2\pi}  \epsilon_{0,{\bm k}} \!\ \sum_{a=1,\cdots} \theta^{<}_a({\bm k}) \theta^{<}_a(-{\bm k}), \nonumber 
\end{align}
\begin{align}
S^{>}_{0} &= \int_{\epsilon_{vN,\Lambda}e^{-2dl}<\epsilon_{vN,{\bm k}}< \epsilon_{vN,\Lambda}} 
\frac{d^2{\bm k}}{2\pi}  \epsilon_{vN,{\bm k}} \,\ \theta^{>}_0({\bm k}) \theta^{>}_0(-{\bm k}) \nonumber \\
& \hspace{-0.7cm} + \int_{\epsilon_{0,\Lambda}e^{-2dl}<\epsilon_{0,{\bm k}}<\epsilon_{0,\Lambda} } 
\frac{d^2{\bm k}}{2\pi}  \epsilon_{0,{\bm k}} \!\ \sum_{a=1,\cdots} \theta^{>}_a({\bm k}) \theta^{>}_a(-{\bm k}), \nonumber 
\end{align}
with RG scale parameter $dl$ in momentum space. Here, we introduce the UV cutoff $\Lambda$ in momentum space while distinguishing the fast and soft modes according to their kinetic energies~\cite{zhao2024}.  
\begin{align}
&\epsilon_{vN,{\bm k}} \equiv \frac{k^2_x}{\sigma_{y}+vNc_y} + \frac{k^2_y}{\sigma_{x}+vNc_x}. \label{epsilon_vN_k_eq}
\end{align}
Namely, the energy of the soft u(1) mode $\theta^{<}_{0}$ is within $[0,\epsilon_{vN,\Lambda} e^{-2dl}]$, and the energy of the soft su$(vN)$ modes $\theta^{<}_{a}$ ($a=1,2,\cdots$) is within $[0,\epsilon_{0,\Lambda} e^{-2dl}]$; the energy of the fast u(1) mode $\theta^{>}_{0}$ is within $[\epsilon_{vN,\Lambda} e^{-2dl},\epsilon_{vN,\Lambda}]$, and the energy of the fast su$(vN)$ modes $\theta^{>}_{a}$ ($a=1,2,\cdots$) is within $[\epsilon_{0,\Lambda} e^{-2dl},\epsilon_{0,\Lambda}]$; the UV energy cutoff is given by, 
\begin{align}
&\epsilon_{vN,\Lambda} \equiv \frac{\Lambda^2}{\sigma_{y}+vNc_y} + \frac{\Lambda^2}{\sigma_{x}+vNc_x}. \label{epsilon_vN_Lambda_eq}
\end{align}
By treating the vortex fugacity $y$ perturbatively, the fast mode is integrated with the free theory $S^{>}_0$:  $\langle ...\rangle_{>} \equiv \frac{1}{Z_{0,>}}\int {\cal D}\Theta^> 
e^{-S^>_{0}} ...$.  Up to the second order in $y$, the integration leads to an effective action for the slow mode, $S^{<}_0 + \langle S_y\rangle_{>} + \frac{1}{2}(\langle S^2_y\rangle_{>}-\langle S_y\rangle^2_{>})$. The first order term $\langle S_y\rangle_{>}$ renormalizes the vortex fugacity $y$ in the effective action for the slow mode,
\begin{align}
y \rightarrow \overline{y} = y \exp\Big[-\frac{1}{2}G({\bm R}={\bm 0})\Big]. \label{y_eq}
\end{align}
Here, a correlation function of the fast mode $G({\bm R})$ is given by the modified Bessel function $K_{1}(x)$ of the second kind~\cite{zhao2024},
\begin{align}
G({\bm R}) &\equiv \frac{v}{2} \Big[g_0({\bm R}) + \frac{g_{vN}({\bm R})-g_0({\bm R})}{N}\Big] dl, \nonumber \\
g_{vN}({\bm r}) & \equiv \sqrt{(\sigma_x+vNc_x)(\sigma_y+vNc_y)} \nonumber \\
&\hspace{1cm} \times \sqrt{\epsilon_{vN,\Lambda}} \!\ r_{vN} K_1(\sqrt{\epsilon_{vN,\Lambda}} \!\ r_{vN}),  
\end{align}
with ${\bm r}_{vN} \equiv (\sqrt{\sigma_y+vNc_y} r_x,\sqrt{\sigma_x+vNc_x} r_y)$, and  $r_{vN} \equiv |{\bm r}_{vN}|$.

$\langle S^2_y\rangle_{>}-\langle S_y\rangle^2_{>}$  renormalizes  conductivity, the Gade constant, and weak topological term in the effective action for the slow mode.  In terms of a gradient expansion for the slowly varying field $\Theta^{<}({\bm r}) = \sum_{a} \theta^{<}_a({\bm r})T_a$,  the relevant terms in $\langle S^2_y\rangle_{>}-\langle S_y\rangle^2_{>}$ are kept up to the second order in the gradient expansion,  
\begin{align}
&\langle S^2_y\rangle_{>}-\langle S_y\rangle^2_{>} =  2y^2 \Lambda^4 \int d^2{\bm r}  
\int d^2 {\bm R} \int d|p\rangle \nonumber \\
& \hspace{1cm} 
\times G({\bm R}) \cos\Big(2\pi v R_{\mu} \langle p | \frac{\partial \Theta^{<}}{\partial r_{\mu}} | p \rangle 
+ \frac{v\chi_y}{4} R_x\Big) +\cdots \nonumber \\
&= 2y^2 \Lambda^4 \int d^2{\bm r} \int d^2 {\bm R} \int d|p\rangle \!\ 
G({\bm R}) \cos\Big(\frac{v\chi_y}{4} R_x\Big) \nonumber \\
&\ \ \ \ \times \bigg\{ 1 - \frac{1}{2} 4 \pi^2 v^2 R^2_{\mu} \langle p | \frac{\partial \Theta^{<}}{\partial r_{\mu}} | p \rangle  \langle p | \frac{\partial \Theta^{<}}{\partial r_{\mu}} | p \rangle  + \cdots \bigg\} \nonumber \\
& \ \ \ + 2y^2 \Lambda^4 \int d^2{\bm r} \int d^2 {\bm R} \int d|p\rangle \!\ 
G({\bm R})  \nonumber \\
&\ \ \ \ \  \times \sin\Big(\frac{v\chi_y}{4} R_x\Big) 
\bigg\{  2\pi v R_{\mu} \langle p | \frac{\partial \Theta^{<}}{\partial r_{\mu}} | p \rangle + \cdots\bigg\} + \cdots. 
\end{align}
Using formula of the integral over $|p\rangle$,    
\begin{align}
&\int d|p\rangle \!\ \langle p|\hat{A}|p\rangle \langle p |\hat{B}|p \rangle =  \nonumber \\
& \left\{\begin{array}{ll}
\frac{\pi^{N-1}}{\Gamma(N+2)} \big({\rm Tr}[\hat{A}] {\rm Tr}[\hat{B}] +
{\rm Tr}[\hat{A}\hat{B}]\big) 
& {\rm for} \,\ {\rm AIII}, \\ 
\frac{1}{4}\frac{\pi^{N/2}}{\Gamma(N/2+2)} \big({\rm Tr}[\hat{A}] {\rm Tr}[\hat{B}] +
2{\rm Tr}[\hat{A}\hat{B}]\big)  & {\rm for} \,\ 
{\rm CII},  \\
 \frac{\pi^{2N-2}}{\Gamma(2N+2)} \big({\rm Tr}[\hat{A}] {\rm Tr}[\hat{B}] +
{\rm Tr}[\hat{A}\hat{B}]\big) & {\rm for}\,\ {\rm BDI}, \\ 
\end{array}\right. 
\end{align}
for Hermitian, real symmetric, and quaternionic self-dual matrices $\hat{A}$ and $\hat{B}$ for the classes AIII, CII, and BDI, respectively, one can rewrite the second-order gradient expansion terms into a renormalization of conductivities and Gade constants, e.g.  
\begin{align}
\delta \Big(\frac{1}{\sigma_y+vNc_y}\Big) = \left\{\begin{array}{ll}
\frac{\pi^N}{\Gamma(N+1)}\Lambda^ 4 y^2 \!\ {\cal B}_x dl & {\rm for} \,\ {\rm AIII},\\
 \frac{\pi^{N/2+1}}{2\Gamma(N/2+1)}\Lambda^ 4 y^2 \!\ {\cal B}_x dl& {\rm for} \,\ {\rm CII},\\
\frac{4\pi^{2N-1}}{\Gamma(2N+1)}\Lambda^ 4 y^2 \!\ {\cal B}_x dl & {\rm for} \,\ {\rm BDI},\\ 
\end{array}\right.
\end{align}
and 
\begin{align}
\delta \Big(\frac{1}{\sigma_y}\Big) = \left\{\begin{array}{ll}
\frac{\pi^N}{\Gamma(N+2)}\Lambda^ 4 y^2 \!\ {\cal B}_x dl & {\rm for} \,\ {\rm AIII},\\
 \frac{\pi^{N/2+1}}{2\Gamma(N/2+2)}\Lambda^ 4 y^2 \!\ {\cal B}_x dl& {\rm for} \,\ {\rm CII},\\
\frac{4\pi^{2N-1}}{\Gamma(2N+2)}\Lambda^ 4 y^2 \!\ {\cal B}_x dl & {\rm for} \,\ {\rm BDI}.\\ 
\end{array}\right.
\end{align}
Here ${\cal B}_{\mu}$ $(\mu=x,y)$ is given by the fast-mode correlation function,
\begin{align}
{\cal B}_{\mu} = &\frac{v}{2}  \int d^2{\bm R} \!\ R^2_{\mu} \cos\Big(\frac{v\chi_y}{4} R_x\Big) \nonumber \\
&\Big( g_0({\bm R}) + \frac{g_{vN}({\bm R})-g_0({\bm R})}{N}\Big). 
\end{align}
 The same form applies for $\delta(1/(\sigma_x+vNc_x))$ and $\delta(1/\sigma_x)$, where ${\cal B}_x$ is replaced by ${\cal B}_y$. Conversely, the first order gradient term remains finite only for the u(1) component of the slow mode, 
\begin{align}
\int d|p\rangle \langle p| \partial_{\mu}\Theta^{<} |p\rangle = \frac{\partial_{\mu}\theta^{<}_0}{\sqrt{vN}}  \left\{\begin{array}{ll} 
\frac{\pi^{N-1}}{\Gamma(N)} & {\rm for}\,\ {\rm AIII}, \\
 \frac{\pi^{N/2}}{\Gamma(N/2)} & {\rm for}\,\ {\rm CII}, \\
 \frac{\pi^{2N-2}}{\Gamma(2N)} & {\rm for}\,\ {\rm BDI}. \\
 \end{array}\right. 
\end{align}
 Thus, by the gauge transformation for $\theta^{<}_0 \rightarrow \theta^{<}_0+\frac{\sqrt{vN}}{8\pi} \delta\chi_y r_x$, one can include the first-order gradient term into a renormalization $\delta \chi_y$ of the weak topological term,   
\begin{align}
\delta \chi_y =-\left\{\begin{array}{ll} 
\frac{4 (\sigma_y+Nc_y)\pi^N}{\Gamma(N+1)} \Lambda^4 y^2 \!\  {\cal C}_x dl, & {\rm for} \,\ {\rm AIII}, \\
\frac{2 (\sigma_y+Nc_y)\pi^{\frac{N}{2}+1}}{\Gamma(N/2+1)} \Lambda^4 y^2 \!\  {\cal C}_x dl, & {\rm for} \,\ {\rm CII}, \\
\frac{8 (\sigma_y+2Nc_y)\pi^{2N-1}}{\Gamma(2N+1)} \Lambda^4 y^2 \!\  {\cal C}_x dl, & {\rm for} \,\ {\rm BDI}, \\
\end{array}\right.
\end{align}
with 
\begin{align}
{\cal C}_x = &\frac{v}{2}  \int d^2{\bm R} \!\ R_x \sin\Big(\frac{v\chi_y}{4} R_x\Big) \nonumber \\
&\Big(
g_0({\bm R}) + \frac{g_{vN}({\bm R})-g_0({\bm R})}{N}\Big). 
\end{align}
Ref.~\cite{zhao2024} did not include this one-loop renormalization $\delta \chi_y$ to the weak topological term, while $\delta \chi_y$ crucially changes the phase diagram [see the main text].  

    A recursive integration of the fast mode followed by an isotropic length rescaling ${\bm r}\rightarrow {\bm r}^{\prime}={\bm r}e^{-dl}$ gives a set of the one-loop RG equations for general $N$. For the chiral unitary (AIII) class,       
\begin{align}
\left\{\begin{array}{l}
\frac{d(\sigma_x+Nc_x)^{-1}}{dl} = \frac{\pi^{N}}{\Gamma(N+1)} \Lambda^4 y^2 \!\ {\cal B}_y,  \\  
\frac{d(\sigma_y+Nc_y)^{-1}}{dl} = \frac{\pi^{N}}{\Gamma(N+1)} \Lambda^4 y^2 \!\ {\cal B}_x,  \\
 \frac{d \sigma^{-1}_x}{dl} = \frac{\pi^{N}}{\Gamma(N+2)} \Lambda^4 y^2 \!\ {\cal B}_y,  \\
 \frac{d \sigma^{-1}_y}{dl}  = \frac{\pi^{N}}{\Gamma(N+2)} \Lambda^4 y^2 \!\ {\cal B}_x, \\ 
\frac{d\chi_y}{dl}  = \chi_y -  \frac{4(\sigma_y + Nc_y) \!\ \pi^{N}}{\Gamma(N+1)} \Lambda^4 y^2 \!\ {\cal C}_x, \\ 
\frac{dy}{dl} = \Big[2 -\big(\frac{\sqrt{\sigma_x\sigma_y}}{4} + \frac{ \sqrt{(\sigma_x+Nc_x)(\sigma_y+Nc_y)}-\sqrt{\sigma_x\sigma_y}}{4N} \big)\Big] y. 
\end{array}\right. \label{rg_aiii}
\end{align}
For the chiral symplectic (CII) class,  
\begin{align}
\left\{\begin{array}{l}
\frac{d(\sigma_x+Nc_x)^{-1}}{dl} = \frac{\pi^{N/2+1}}{2\Gamma(N/2+1)} \Lambda^4 y^2 \!\ {\cal B}_y,  \\  
\frac{d(\sigma_y+Nc_y)^{-1}}{dl} = \frac{\pi^{N/2+1}}{2\Gamma(N/2+1)} \Lambda^4 y^2 \!\ {\cal B}_x,  \\
 \frac{d \sigma^{-1}_x}{dl} = \frac{\pi^{N/2+1}}{2\Gamma(N/2+2)} \Lambda^4 y^2 \!\ {\cal B}_y,  \\
 \frac{d \sigma^{-1}_y}{dl}  = \frac{\pi^{N/2+1}}{2\Gamma(N/2+2)} \Lambda^4 y^2 \!\ {\cal B}_x, \\ 
\frac{d\chi_y}{dl}  = \chi_y -  \frac{2(\sigma_y + Nc_y) \!\ \pi^{N/2+1}}{\Gamma(N/2+1)} \Lambda^4 y^2 \!\ {\cal C}_x, \\ 
\frac{dy}{dl} = \Big[2 -\big(\frac{\sqrt{\sigma_x\sigma_y}}{4} + \frac{ \sqrt{(\sigma_x+Nc_x)(\sigma_y+Nc_y)}-\sqrt{\sigma_x\sigma_y}}{4N} \big)\Big] y. 
\end{array}\right. \label{rg_cii}
\end{align}
For the chiral orthogonal (BDI) class,  
\begin{align}
\left\{\begin{array}{l}
\frac{d(\sigma_x+2Nc_x)^{-1}}{dl} = \frac{4\pi^{2N-1}}{\Gamma(2N+1)} \Lambda^4 y^2 \!\ {\cal B}_y,  \\  
\frac{d(\sigma_y+2Nc_y)^{-1}}{dl} = \frac{4\pi^{2N-1}}{\Gamma(2N+1)} \Lambda^4 y^2 \!\ {\cal B}_x,  \\
 \frac{d \sigma^{-1}_x}{dl} = \frac{4\pi^{2N-1}}{\Gamma(2N+2)} \Lambda^4 y^2 \!\ {\cal B}_y,  \\
 \frac{d \sigma^{-1}_y}{dl}  = \frac{4\pi^{2N-1}}{\Gamma(2N+2)} \Lambda^4 y^2 \!\ {\cal B}_x, \\ 
\frac{d\chi_y}{dl}  = \chi_y -  \frac{8(\sigma_y + 2Nc_y) \!\ \pi^{2N-1}}{\Gamma(2N+1)} \Lambda^4 y^2 \!\ {\cal C}_x, \\ 
\frac{dy}{dl} = \Big[2 -\big(\frac{\sqrt{\sigma_x\sigma_y}}{2} + \frac{ \sqrt{(\sigma_x+2Nc_x)(\sigma_y+2Nc_y)}-\sqrt{\sigma_x\sigma_y}}{2N} \big)\Big] y. 
\end{array}\right. \label{rg_bdi}
\end{align}

Here ${\cal B}_y$, ${\cal B}_x$ and ${\cal C}_x$ are functions of the coupling constants, and they are given by the followings,   
\begin{align}
\left\{\begin{array}{l}
 {\cal B}_\mu =  \frac{v}{2} \Big[I_{0,\mu}+\frac{I_{vN,\mu}-I_{0,\mu}}{N}\Big], \,\ \,\ {\rm for}\,\ \,\ \mu=x,y,  \\
 {\cal C}_x = \frac{v}{2}\Big[L_{0,x} + \frac{L_{vN,x}-L_{0,x}}{N}\Big], \\
 \end{array}\right. 
\end{align}
and 
\begin{align}
\left\{\begin{array}{l}
I_{vN,x} = \frac{16\pi}{\epsilon^2_{vN,\Lambda}(\sigma_y+vNc_y)}\frac{1-5\tilde{\chi}^2_{vN}}{(1+\tilde{\chi}^2_{vN})^4},  \\
I_{vN,y} = \frac{16\pi}{\epsilon^2_{vN,\Lambda}(\sigma_x+vNc_x)}\frac{1}{(1+\tilde{\chi}^2_{vN})^3},  \\
L_{vN,x} = \frac{16\pi}{\epsilon^{\frac{3}{2}}_{vN,\Lambda}(\sigma_y+vNc_y)^{\frac{1}{2}}}\frac{\tilde{\chi}_{vN}}{(1+\tilde{\chi}^2_{vN})^3},  \\ 
\tilde{\chi}_{vN} = \frac{v\chi_y}{4\sqrt{\epsilon_{vN,\Lambda}(\sigma_y+vNc_y)}}, 
\end{array}\right.
\end{align}
with $v=1$ for AIII and CII, $v=2$ for BDI. In the replica limit $(N\rightarrow 0)$, Eqs.~(\ref{rg_aiii},\ref{rg_cii},\ref{rg_bdi}) reduce to Eqs.~(\ref{a3},\ref{c2},\ref{bd1}), respectively.

\bibliography{ref}

@misc{shang2025,
  title={Anisotropic Anderson localization in higher-dimensional nonreciprocal lattices},
  author={Jinyuan Shang and Haiping Hu},
  year={2025},
  eprint={2507.14523},
  archivePrefix={arXiv},
  primaryClass={cond-mat.dis-nn},
  url={https://arxiv.org/abs/2507.14523},
}

@article{luo2018b,
  title = {Unconventional scaling theory in disorder-driven quantum phase transition},
  author = {Luo, Xunlong and Ohtsuki, Tomi and Shindou, Ryuichi},
  journal = {Phys. Rev. B},
  volume = {98},
  issue = {2},
  pages = {020201(R)},
  numpages = {5},
  year = {2018},
  month = {Jul},
  publisher = {American Physical Society},
  doi = {10.1103/PhysRevB.98.020201},
  url = {https://link.aps.org/doi/10.1103/PhysRevB.98.020201}
}

@article{luo2018a,
  title = {Quantum multicriticality in disordered Weyl semimetals},
  author = {Luo, Xunlong and Xu, Baolong and Ohtsuki, Tomi and Shindou, Ryuichi},
  journal = {Phys. Rev. B},
  volume = {97},
  issue = {4},
  pages = {045129},
  numpages = {21},
  year = {2018},
  month = {Jan},
  publisher = {American Physical Society},
  doi = {10.1103/PhysRevB.97.045129},
  url = {https://link.aps.org/doi/10.1103/PhysRevB.97.045129}
}

@article{zhao2024,
  title = {Topological Effect on the Anderson Transition in Chiral Symmetry Classes},
  author = {Zhao, Pengwei and Xiao, Zhenyu and Zhang, Yeyang and Shindou, Ryuichi},
  journal = {Phys. Rev. Lett.},
  volume = {133},
  issue = {22},
  pages = {226601},
  numpages = {7},
  year = {2024},
  month = {Nov},
  publisher = {American Physical Society},
  doi = {10.1103/PhysRevLett.133.226601},
  url = {https://link.aps.org/doi/10.1103/PhysRevLett.133.226601}
}

@misc{zhao2025,
  title={Theory of the Anderson transition in three-dimensional chiral symmetry classes: Connection to type-II superconductors},
  author={Pengwei Zhao and Ryuichi Shindou},
  year={2025},
  eprint={2506.21050},
  archivePrefix={arXiv},
  primaryClass={cond-mat.dis-nn},
  url={https://arxiv.org/abs/2506.21050},
}

@article{xiao2023,
  title = {Anisotropic Topological Anderson Transitions in Chiral Symmetry Classes},
  author = {Xiao, Zhenyu and Kawabata, Kohei and Luo, Xunlong and Ohtsuki, Tomi and Shindou, Ryuichi},
  journal = {Phys. Rev. Lett.},
  volume = {131},
  issue = {5},
  pages = {056301},
  numpages = {8},
  year = {2023},
  month = {Aug},
  publisher = {American Physical Society},
  doi = {10.1103/PhysRevLett.131.056301},
  url = {https://link.aps.org/doi/10.1103/PhysRevLett.131.056301}
}

@article{asada2002,
  title = {Anderson Transition in Two-Dimensional Systems with Spin-Orbit Coupling},
  author = {Asada, Yoichi and Slevin, Keith and Ohtsuki, Tomi},
  journal = {Phys. Rev. Lett.},
  volume = {89},
  issue = {25},
  pages = {256601},
  numpages = {4},
  year = {2002},
  month = {Dec},
  publisher = {American Physical Society},
  doi = {10.1103/PhysRevLett.89.256601},
  url = {https://link.aps.org/doi/10.1103/PhysRevLett.89.256601}
}

@article{luo2022,
  title = {Unifying the Anderson transitions in Hermitian and non-Hermitian systems},
  author = {Luo, Xunlong and Xiao, Zhenyu and Kawabata, Kohei and Ohtsuki, Tomi and Shindou, Ryuichi},
  journal = {Phys. Rev. Res.},
  volume = {4},
  issue = {2},
  pages = {L022035},
  numpages = {7},
  year = {2022},
  month = {May},
  publisher = {American Physical Society},
  doi = {10.1103/PhysRevResearch.4.L022035},
  url = {https://link.aps.org/doi/10.1103/PhysRevResearch.4.L022035}
}

@article{wang2021a,
  title = {Universality classes of the Anderson transition in the three-dimensional symmetry classes AIII, BDI, C, D, and CI},
  author = {Wang, Tong and Ohtsuki, Tomi and Shindou, Ryuichi},
  journal = {Phys. Rev. B},
  volume = {104},
  issue = {1},
  pages = {014206},
  numpages = {10},
  year = {2021},
  month = {Jul},
  publisher = {American Physical Society},
  doi = {10.1103/PhysRevB.104.014206},
  url = {https://link.aps.org/doi/10.1103/PhysRevB.104.014206}
}

@article{slevin2014,
  doi = {10.1088/1367-2630/16/1/015012},
  url = {https://dx.doi.org/10.1088/1367-2630/16/1/015012},
  year = {2014},
  month = {jan},
  publisher = {IOP Publishing},
  volume = {16},
  number = {1},
  pages = {015012},
  author = {Slevin, Keith and Ohtsuki, Tomi},
  title = {Critical exponent for the Anderson transition in the three-dimensional orthogonal universality class},
  journal = {New Journal of Physics}
}

@article{evers2008,
  title = {Anderson transitions},
  author = {Evers, Ferdinand and Mirlin, Alexander D.},
  journal = {Rev. Mod. Phys.},
  volume = {80},
  issue = {4},
  pages = {1355--1417},
  numpages = {0},
  year = {2008},
  month = {Oct},
  publisher = {American Physical Society},
  doi = {10.1103/RevModPhys.80.1355},
  url = {https://link.aps.org/doi/10.1103/RevModPhys.80.1355}
}

@article{kosterlitz1972,
  doi = {10.1088/0022-3719/5/11/002},
  url = {https://dx.doi.org/10.1088/0022-3719/5/11/002},
  year = {1972},
  month = {jun},
  publisher = {},
  volume = {5},
  number = {11},
  pages = {L124},
  author = {J M Kosterlitz and D J Thouless},
  title = {Long range order and metastability in two dimensional solids and superfluids. (Application of dislocation theory)},
  journal = {Journal of Physics C: Solid State Physics}
}

@article{kosterlitz1973,
  doi = {10.1088/0022-3719/6/7/010},
  url = {https://dx.doi.org/10.1088/0022-3719/6/7/010},
  year = {1973},
  month = {apr},
  publisher = {},
  volume = {6},
  number = {7},
  pages = {1181},
  author = {J M Kosterlitz and D J Thouless},
  title = {Ordering, metastability and phase transitions in two-dimensional systems},
  journal = {Journal of Physics C: Solid State Physics}
}

@article{kosterlitz1974,
  doi = {10.1088/0022-3719/7/6/005},
  url = {https://dx.doi.org/10.1088/0022-3719/7/6/005},
  year = {1974},
  month = {mar},
  publisher = {},
  volume = {7},
  number = {6},
  pages = {1046},
  author = {J M Kosterlitz},
  title = {The critical properties of the two-dimensional xy model},
  journal = {Journal of Physics C: Solid State Physics}
}

@article{konig2012,
  title = {Metal-insulator transition in two-dimensional random fermion systems of chiral symmetry classes},
  author = {K\"onig, E. J. and Ostrovsky, P. M. and Protopopov, I. V. and Mirlin, A. D.},
  journal = {Phys. Rev. B},
  volume = {85},
  issue = {19},
  pages = {195130},
  numpages = {18},
  year = {2012},
  month = {May},
  publisher = {American Physical Society},
  doi = {10.1103/PhysRevB.85.195130},
  url = {https://link.aps.org/doi/10.1103/PhysRevB.85.195130}
}

@article{mackinnon1981,
  title = {One-Parameter Scaling of Localization Length and Conductance in Disordered Systems},
  author = {MacKinnon, A. and Kramer, B.},
  journal = {Phys. Rev. Lett.},
  volume = {47},
  issue = {21},
  pages = {1546--1549},
  numpages = {0},
  year = {1981},
  month = {Nov},
  publisher = {American Physical Society},
  doi = {10.1103/PhysRevLett.47.1546},
  url = {https://link.aps.org/doi/10.1103/PhysRevLett.47.1546}
}

@article{mackinnon1983,
  title={The scaling theory of electrons in disordered solids: Additional numerical results},
  author={MacKinnon, A and Kramer, B},
  journal={Zeitschrift f{\"u}r Physik B Condensed Matter},
  volume={53},
  number={1},
  pages={1--13},
  year={1983},
  publisher={Springer}
}

@article{pichard1981,
  doi = {10.1088/0022-3719/14/6/003},
  url = {https://dx.doi.org/10.1088/0022-3719/14/6/003},
  year = {1981},
  month = {feb},
  publisher = {},
  volume = {14},
  number = {6},
  pages = {L127},
  author = {J L Pichard and G Sarma},
  title = {Finite size scaling approach to Anderson localisation},
  journal = {Journal of Physics C: Solid State Physics}
}

@article{slevin1999,
  title = {Corrections to Scaling at the Anderson Transition},
  author = {Slevin, Keith and Ohtsuki, Tomi},
  journal = {Phys. Rev. Lett.},
  volume = {82},
  issue = {2},
  pages = {382--385},
  numpages = {0},
  year = {1999},
  month = {Jan},
  publisher = {American Physical Society},
  doi = {10.1103/PhysRevLett.82.382},
  url = {https://link.aps.org/doi/10.1103/PhysRevLett.82.382}
}

@book{crisanti2012,
  title={Products of random matrices: in Statistical Physics},
  author={Crisanti, Andrea and Paladin, Giovanni and Vulpiani, Angelo},
  volume={104},
  year={2012},
  publisher={Springer Science \& Business Media}
}

@book{cardy1996,
  title={Scaling and renormalization in statistical physics},
  author={Cardy, John},
  volume={5},
  year={1996},
  publisher={Cambridge university press}
}

@article{mondragon2014,
  title = {Topological Criticality in the Chiral-Symmetric AIII Class at Strong Disorder},
  author = {Mondragon-Shem, Ian and Hughes, Taylor L. and Song, Juntao and Prodan, Emil},
  journal = {Phys. Rev. Lett.},
  volume = {113},
  issue = {4},
  pages = {046802},
  numpages = {5},
  year = {2014},
  month = {Jul},
  publisher = {American Physical Society},
  doi = {10.1103/PhysRevLett.113.046802},
  url = {https://link.aps.org/doi/10.1103/PhysRevLett.113.046802}
}

@article{altland2014,
  title = {Quantum Criticality of Quasi-One-Dimensional Topological Anderson Insulators},
  author = {Altland, Alexander and Bagrets, Dmitry and Fritz, Lars and Kamenev, Alex and Schmiedt, Hanno},
  journal = {Phys. Rev. Lett.},
  volume = {112},
  issue = {20},
  pages = {206602},
  numpages = {5},
  year = {2014},
  month = {May},
  publisher = {American Physical Society},
  doi = {10.1103/PhysRevLett.112.206602},
  url = {https://link.aps.org/doi/10.1103/PhysRevLett.112.206602}
}

@article{claes2020,
  title = {Disorder driven phase transitions in weak AIII topological insulators},
  author = {Claes, Jahan and Hughes, Taylor L.},
  journal = {Phys. Rev. B},
  volume = {101},
  issue = {22},
  pages = {224201},
  numpages = {7},
  year = {2020},
  month = {Jun},
  publisher = {American Physical Society},
  doi = {10.1103/PhysRevB.101.224201},
  url = {https://link.aps.org/doi/10.1103/PhysRevB.101.224201}
}

@article{luca2003,
  doi = {10.1088/0305-4470/36/14/311},
  url = {https://dx.doi.org/10.1088/0305-4470/36/14/311},
  year = {2003},
  month = {mar},
  publisher = {},
  volume = {36},
  number = {14},
  pages = {4081},
  author = {Luca Molinari},
  title = {Spectral duality and distribution of exponents for transfer matrices of block-tridiagonal Hamiltonians},
  journal = {Journal of Physics A: Mathematical and General}
}

@article{hatano1996,
  title = {Localization Transitions in Non-Hermitian Quantum Mechanics},
  author = {Hatano, Naomichi and Nelson, David R.},
  journal = {Phys. Rev. Lett.},
  volume = {77},
  issue = {3},
  pages = {570--573},
  numpages = {0},
  year = {1996},
  month = {Jul},
  publisher = {American Physical Society},
  doi = {10.1103/PhysRevLett.77.570},
  url = {https://link.aps.org/doi/10.1103/PhysRevLett.77.570}
}

@article{asada2004,
  title = {Numerical estimation of the $\ensuremath{\beta}$ function in two-dimensional systems with spin-orbit coupling},
  author = {Asada, Yoichi and Slevin, Keith and Ohtsuki, Tomi},
  journal = {Phys. Rev. B},
  volume = {70},
  issue = {3},
  pages = {035115},
  numpages = {6},
  year = {2004},
  month = {Jul},
  publisher = {American Physical Society},
  doi = {10.1103/PhysRevB.70.035115},
  url = {https://link.aps.org/doi/10.1103/PhysRevB.70.035115}
}

@article{li2020,
  title = {Topological Phase Transitions in Disordered Electric Quadrupole Insulators},
  author = {Li, Chang-An and Fu, Bo and Hu, Zi-Ang and Li, Jian and Shen, Shun-Qing},
  journal = {Phys. Rev. Lett.},
  volume = {125},
  issue = {16},
  pages = {166801},
  numpages = {7},
  year = {2020},
  month = {Oct},
  publisher = {American Physical Society},
  doi = {10.1103/PhysRevLett.125.166801},
  url = {https://link.aps.org/doi/10.1103/PhysRevLett.125.166801}
}

@article{silva2025,
  title = {Disorder-induced instability of a Weyl nodal loop semimetal towards a diffusive topological metal with protected multifractal surface states},
  author = {Silva, Jo\~ao S. and Gon\ifmmode \mbox{\c{c}}\else \c{c}\fi{}alves, Miguel and Castro, Eduardo V. and Ribeiro, Pedro and Ara\'ujo, Miguel A. N.},
  journal = {Phys. Rev. B},
  volume = {111},
  issue = {4},
  pages = {L041116},
  numpages = {8},
  year = {2025},
  month = {Jan},
  publisher = {American Physical Society},
  doi = {10.1103/PhysRevB.111.L041116},
  url = {https://link.aps.org/doi/10.1103/PhysRevB.111.L041116}
}

@article{garcia-garcia2006,
  title = {Anderson transition in systems with chiral symmetry},
  author = {Garc\'{\i}a-Garc\'{\i}a, Antonio M. and Cuevas, Emilio},
  journal = {Phys. Rev. B},
  volume = {74},
  issue = {11},
  pages = {113101},
  numpages = {4},
  year = {2006},
  month = {Sep},
  publisher = {American Physical Society},
  doi = {10.1103/PhysRevB.74.113101},
  url = {https://link.aps.org/doi/10.1103/PhysRevB.74.113101}
}

@article{onsager1949,
  author = {Lars Onsager},
  title = {Statistical Hydrodynamics},
  journal = {Nuovo Cimento (Supplemento)},
  volume = {6},
  pages = {279-287},
  year = {1949}
}

@article{feynman1955,
  author = {Richard P. Feynman},
  title = {Application of Quantum Mechanics to Liquid Helium},
  journal = {Progress in Low Temperature Physics},
  volume = {1},
  pages = {17-53},
  year = {1955}
}

@article{berezinskii71,
  url = {https://jetp.ras.ru/cgi-bin/dn/e_032_03_0493.pdf},
  year = {1971},
  volume = {32},
  number = {3},
  pages = {493},
  author = {V L Berezinskii},
  title = {Destruction of Long-range Order in One-dimensional and Two-dimensional Systems having a Continuous Symmetry Group I. Classical Systems},
  journal = {Soviet Journal of Experimental and Theoretical Physics},
}

@article{berezinskii72,
  doi = {},
  url = {https://jetp.ras.ru/cgi-bin/dn/e_032_03_0493.pdf},
  year = {1972},
  month = {},
  publisher = {},
  volume = {34},
  number = {3},
  pages = {610},
  author = {V L Berezinskii},
  title = {Destruction of Long-range Order in One-dimensional and Two-dimensional Systems having a Continuous Symmetry Group II. Quantum  Systems},
  journal = {Soviet Journal of Experimental and Theoretical Physics}
}

@article{nelson1977,
  title = {Dynamics of classical $\mathrm{XY}$ spins in one and two dimensions},
  author = {Nelson, David R. and Fisher, Daniel S.},
  journal = {Phys. Rev. B},
  volume = {16},
  issue = {11},
  pages = {4945--4955},
  numpages = {0},
  year = {1977},
  month = {Dec},
  publisher = {American Physical Society},
  doi = {10.1103/PhysRevB.16.4945},
  url = {https://link.aps.org/doi/10.1103/PhysRevB.16.4945}
}

@article{dasgupta1981,
  title = {Phase Transition in a Lattice Model of Superconductivity},
  author = {Dasgupta, C. and Halperin, B. I.},
  journal = {Phys. Rev. Lett.},
  volume = {47},
  issue = {21},
  pages = {1556--1560},
  numpages = {0},
  year = {1981},
  month = {Nov},
  publisher = {American Physical Society},
  doi = {10.1103/PhysRevLett.47.1556},
  url = {https://link.aps.org/doi/10.1103/PhysRevLett.47.1556}
}

@article{williams1987,
  title = {Vortex-ring model of the superfluid \ensuremath{\lambda} transition},
  author = {Williams, Gary A.},
  journal = {Phys. Rev. Lett.},
  volume = {59},
  issue = {17},
  pages = {1926--1929},
  numpages = {0},
  year = {1987},
  month = {Oct},
  publisher = {American Physical Society},
  doi = {10.1103/PhysRevLett.59.1926},
  url = {https://link.aps.org/doi/10.1103/PhysRevLett.59.1926}
}

@article{shenoy1989,
  title = {Vortex-loop scaling in the three-dimensional XY ferromagnet},
  author = {Shenoy, Subodh R.},
  journal = {Phys. Rev. B},
  volume = {40},
  issue = {7},
  pages = {5056--5068},
  numpages = {0},
  year = {1989},
  month = {Sep},
  publisher = {American Physical Society},
  doi = {10.1103/PhysRevB.40.5056},
  url = {https://link.aps.org/doi/10.1103/PhysRevB.40.5056}
}

@article{shindou2025,
  title = {Topological effect on order-disorder transitions in U(1) sigma models},
  author = {Shindou, Ryuichi and Zhao, Pengwei},
  journal = {Phys. Rev. B},
  volume = {111},
  issue = {22},
  pages = {224507},
  numpages = {27},
  year = {2025},
  month = {Jun},
  publisher = {American Physical Society},
  doi = {10.1103/hydl-dyxj},
  url = {https://link.aps.org/doi/10.1103/hydl-dyxj}
}

@article{abrahams1979,
  title = {Scaling Theory of Localization: Absence of Quantum Diffusion in Two Dimensions},
  author = {Abrahams, E. and Anderson, P. W. and Licciardello, D. C. and Ramakrishnan, T. V.},
  journal = {Phys. Rev. Lett.},
  volume = {42},
  issue = {10},
  pages = {673--676},
  numpages = {0},
  year = {1979},
  month = {Mar},
  publisher = {American Physical Society},
  doi = {10.1103/PhysRevLett.42.673},
  url = {https://link.aps.org/doi/10.1103/PhysRevLett.42.673}
}

@article{wegner1976,
  title = {Electrons in disordered systems. Scaling near the mobility edge.},
  author = {Wegner, F. J.},
  journal = {Z. Physik B Condensed Matter},
  volume = {25},
  pages = {327-337},
  year = {1976},
  doi = {https://doi.org/10.1007/BF01315248}
}

@article{wegner1979,
  title = {The mobility edge problem: Continuous symmetry and a conjecture.},
  author = {Wegner, F.},
  journal = {Z. Physik B Condensed Matter},
  volume = {35},
  pages = {207-210},
  year = {1979},
  doi = {https://doi.org/10.1007/BF01319839}
}

@article{efetov1980,
  author = {Efetov, K. B. and Larkin, A. I. and Khmel'nitskii, D. E.},
  title = {Interaction of diffusion modes in the theory of localization.},
  journal = {Sov. Phys. JETP},
  volume = {52},
  pages = {568},
  year = {1980}
}

@article{hikami1980,
  author = {Hikami, Shinobu and Larkin, Anatoly I. and Nagaoka, Yosuke},
  title = {Spin-Orbit Interaction and Magnetoresistance in the Two Dimensional Random System},
  journal = {Progress of Theoretical Physics},
  volume = {63},
  number = {2},
  pages = {707-710},
  year = {1980},
  month = {02},
  issn = {0033-068X},
  doi = {10.1143/PTP.63.707},
  url = {https://doi.org/10.1143/PTP.63.707},
  eprint = {https://academic.oup.com/ptp/article-pdf/63/2/707/5336056/63-2-707.pdf},
}

@article{hikami1981,
  title = {Anderson localization in a nonlinear-$\ensuremath{\sigma}$-model representation},
  author = {Hikami, Shinobu},
  journal = {Phys. Rev. B},
  volume = {24},
  issue = {5},
  pages = {2671--2679},
  numpages = {0},
  year = {1981},
  month = {Sep},
  publisher = {American Physical Society},
  doi = {10.1103/PhysRevB.24.2671},
  url = {https://link.aps.org/doi/10.1103/PhysRevB.24.2671}
}

@article{efetov1983,
  author = {K. B. Efetov},
  title = {Supersymmetry and theory of disordered metals},
  journal = {Advances in Physics},
  volume = {32},
  number = {1},
  pages = {53--127},
  year = {1983},
  publisher = {Taylor \& Francis},
  doi = {10.1080/00018738300101531},
  URL = {https://doi.org/10.1080/00018738300101531},
  eprint = {https://doi.org/10.1080/00018738300101531}
}

@article{chen2025,
  title = {Field theory of non-Hermitian disordered systems},
  author = {Chen, Ze and Kawabata, Kohei and Kulkarni, Anish and Ryu, Shinsei},
  journal = {Phys. Rev. B},
  volume = {111},
  issue = {5},
  pages = {054203},
  numpages = {22},
  year = {2025},
  month = {Feb},
  publisher = {American Physical Society},
  doi = {10.1103/PhysRevB.111.054203},
  url = {https://link.aps.org/doi/10.1103/PhysRevB.111.054203}
}

@article{altland1999,
  title = {Field Theory of the Random Flux Model},
  author = {Altland, Alexander and Simons, B. D.},
  year = {1999},
  journal = {Nuclear Physics B},
  volume = {562},
  number = {3},
  pages = {445--476},
  issn = {0550-3213},
  doi = {10.1016/S0550-3213(99)00543-X}
}

@article{altland1997,
  title = {Nonstandard Symmetry Classes in Mesoscopic Normal-Superconducting Hybrid Structures},
  author = {Altland, Alexander and Zirnbauer, Martin R.},
  year = {1997},
  journal = {Physical Review B},
  volume = {55},
  number = {2},
  pages = {1142--1161},
  publisher = {American Physical Society},
  doi = {10.1103/PhysRevB.55.1142}
}

@article{altland2015,
  title = {Topology versus {{Anderson}} Localization: {{Nonperturbative}} Solutions in One Dimension},
  author = {Altland, Alexander and Bagrets, Dmitry and Kamenev, Alex},
  year = {2015},
  journal = {Physical Review B},
  volume = {91},
  number = {8},
  pages = {085429},
  publisher = {American Physical Society},
  doi = {10.1103/PhysRevB.91.085429}
}

@article{balents1997,
  title = {Delocalization Transition via Supersymmetry in One Dimension},
  author = {Balents, Leon and Fisher, Matthew P. A.},
  year = {1997},
  journal = {Physical Review B},
  volume = {56},
  number = {20},
  pages = {12970--12991},
  publisher = {American Physical Society},
  doi = {10.1103/PhysRevB.56.12970}
}

@article{fabrizio2000,
  title = {Anderson Localization in Bipartite Lattices},
  author = {Fabrizio, Michele and Castellani, Claudio},
  year = {2000},
  journal = {Nuclear Physics B},
  volume = {583},
  number = {3},
  pages = {542--583},
  issn = {0550-3213},
  doi = {10.1016/S0550-3213(00)00311-4},
  langid = {english}
}

@article{fukui1999,
  title = {Critical Behavior of Two-Dimensional Random Hopping Fermions with {$\pi$}-Flux},
  author = {Fukui, Takahiro},
  year = {1999},
  journal = {Nuclear Physics B},
  volume = {562},
  number = {3},
  pages = {477--496},
  issn = {0550-3213},
  doi = {10.1016/S0550-3213(99)00494-0},
  langid = {english}
}

@article{gade1993,
  title = {Anderson Localization for Sublattice Models},
  author = {Gade, Renate},
  year = {1993},
  journal = {Nuclear Physics B},
  volume = {398},
  number = {3},
  pages = {499--515},
  issn = {0550-3213},
  doi = {10.1016/0550-3213(93)90601-K}
}

@article{gade1991,
  title = {The n = 0 Replica Limit of {{U}}(n) and {{U}}(n){{SO}}(n) Models},
  author = {Gade, Renate and Wegner, Franz},
  year = {1991},
  journal = {Nuclear Physics B},
  volume = {360},
  number = {2},
  pages = {213--218},
  issn = {0550-3213},
  doi = {10.1016/0550-3213(91)90401-I}
}

@book{giamarchi2004,
  title = {Quantum {{Physics}} in {{One Dimension}}},
  author = {Giamarchi, Thierry},
  year = {2004},
  publisher = {Clarendon Press},
  isbn = {978-0-19-852500-4},
  langid = {english}
}

@article{popov1973,
  author = {Popov, V. N.},
  title = {Quantum vortices and phase transitions in Bose systems},
  journal = {Soviet Physics -- JETP},
  volume = {37},
  pages = {341},
  year = {1973}
}

@article{wiegel1973,
  title = {Vortex-ring model of Bose condensation},
  journal = {Physica},
  volume = {65},
  number = {2},
  pages = {321-336},
  year = {1973},
  issn = {0031-8914},
  doi = {https://doi.org/10.1016/0031-8914(73)90348-0},
  url = {https://www.sciencedirect.com/science/article/pii/0031891473903480},
  author = {F.W. Wiegel},
}

@article{karcher2023a,
  title = {Metal-Insulator Transition in a Two-Dimensional System of Chiral Unitary Class},
  author = {Karcher, Jonas F. and Gruzberg, Ilya A. and Mirlin, Alexander D.},
  year = {2023},
  journal = {Physical Review B},
  volume = {107},
  number = {2},
  pages = {L020201},
  publisher = {American Physical Society},
  doi = {10.1103/PhysRevB.107.L020201}
}

@article{karcher2023b,
  title = {Generalized multifractality in two-dimensional disordered systems of chiral symmetry classes},
  author = {Karcher, Jonas F. and Gruzberg, Ilya A. and Mirlin, Alexander D.},
  journal = {Phys. Rev. B},
  volume = {107},
  issue = {10},
  pages = {104202},
  numpages = {25},
  year = {2023},
  month = {Mar},
  publisher = {American Physical Society},
  doi = {10.1103/PhysRevB.107.104202},
  url = {https://link.aps.org/doi/10.1103/PhysRevB.107.104202}
}

@article{kawabata2019a,
  title = {Symmetry and {{Topology}} in {{Non-Hermitian Physics}}},
  author = {Kawabata, Kohei and Shiozaki, Ken and Ueda, Masahito and Sato, Masatoshi},
  year = {2019},
  journal = {Physical Review X},
  volume = {9},
  number = {4},
  pages = {041015},
  publisher = {American Physical Society},
  doi = {10.1103/PhysRevX.9.041015}
}

@article{Ryu2010,
  doi = {10.1088/1367-2630/12/6/065010},
  url = {https://dx.doi.org/10.1088/1367-2630/12/6/065010},
  year = {2010},
  month = {jun},
  publisher = {},
  volume = {12},
  number = {6},
  pages = {065010},
  author = {Ryu, Shinsei and Schnyder, Andreas P and Furusaki, Akira and Ludwig, Andreas W W},
  title = {Topological insulators and superconductors: tenfold way and dimensional hierarchy},
  journal = {New Journal of Physics}
}

@article{li2022,
  title = {Topological {{States}} in {{Two-Dimensional Su-Schrieffer-Heeger Models}}},
  author = {Li, Chang-An},
  year = {2022},
  journal = {Frontiers in Physics},
  volume = {10},
  issn = {2296-424X}
}

@article{luo2020,
  title = {Critical Behavior of {{Anderson}} Transitions in Three-Dimensional Orthogonal Classes with Particle-Hole Symmetries},
  author = {Luo, Xunlong and Xu, Baolong and Ohtsuki, Tomi and Shindou, Ryuichi},
  year = {2020},
  journal = {Physical Review B},
  volume = {101},
  number = {2},
  pages = {020202},
  publisher = {American Physical Society},
  doi = {10.1103/PhysRevB.101.020202}
}

@article{luo2021b,
  title = {Transfer Matrix Study of the {{Anderson}} Transition in Non-{{Hermitian}} Systems},
  author = {Luo, Xunlong and Ohtsuki, Tomi and Shindou, Ryuichi},
  year = {2021},
  journal = {Physical Review B},
  volume = {104},
  number = {10},
  pages = {104203},
  publisher = {American Physical Society},
  doi = {10.1103/PhysRevB.104.104203}
}

@article{tanaka2015,
  title = {A Short Guide to Topological Terms in the Effective Theories of Condensed Matter},
  author = {Tanaka, Akihiro and Takayoshi, Shintaro},
  year = {2015},
  journal = {Science and Technology of Advanced Materials},
  volume = {16},
  number = {1},
  pages = {014404},
  publisher = {IOP Publishing},
  issn = {1468-6996},
  doi = {10.1088/1468-6996/16/1/014404},
  langid = {english}
}

@article{li2025,
  title = {Random-flux-induced transition sequence between weak and strong topological phases with anisotropic localization properties},
  author = {Li, Chang-An and Fu, Bo and Li, Jian and Trauzettel, Bj\"orn},
  journal = {Phys. Rev. B},
  volume = {111},
  issue = {21},
  pages = {214207},
  numpages = {15},
  year = {2025},
  month = {Jun},
  publisher = {American Physical Society},
  doi = {10.1103/zjyw-ln2n},
  url = {https://link.aps.org/doi/10.1103/zjyw-ln2n}
}

@article{schnyder2011,
  title = {Topological phases and surface flat bands in superconductors without inversion symmetry},
  author = {Schnyder, Andreas P. and Ryu, Shinsei},
  journal = {Phys. Rev. B},
  volume = {84},
  issue = {6},
  pages = {060504},
  numpages = {4},
  year = {2011},
  month = {Aug},
  publisher = {American Physical Society},
  doi = {10.1103/PhysRevB.84.060504},
  url = {https://link.aps.org/doi/10.1103/PhysRevB.84.060504}
}

@article{fulga2011,
  title = {Scattering formula for the topological quantum number of a disordered multimode wire},
  author = {Fulga, I. C. and Hassler, F. and Akhmerov, A. R. and Beenakker, C. W. J.},
  journal = {Phys. Rev. B},
  volume = {83},
  issue = {15},
  pages = {155429},
  numpages = {8},
  year = {2011},
  month = {Apr},
  publisher = {American Physical Society},
  doi = {10.1103/PhysRevB.83.155429},
  url = {https://link.aps.org/doi/10.1103/PhysRevB.83.155429}
}

@article{fulga2012a,
  title = {Scattering theory of topological insulators and superconductors},
  author = {Fulga, I. C. and Hassler, F. and Akhmerov, A. R.},
  journal = {Phys. Rev. B},
  volume = {85},
  issue = {16},
  pages = {165409},
  numpages = {12},
  year = {2012},
  month = {Apr},
  publisher = {American Physical Society},
  doi = {10.1103/PhysRevB.85.165409},
  url = {https://link.aps.org/doi/10.1103/PhysRevB.85.165409}
}

@article{esaki2011,
  title = {Edge states and topological phases in non-Hermitian systems},
  author = {Esaki, Kenta and Sato, Masatoshi and Hasebe, Kazuki and Kohmoto, Mahito},
  journal = {Phys. Rev. B},
  volume = {84},
  issue = {20},
  pages = {205128},
  numpages = {19},
  year = {2011},
  month = {Nov},
  publisher = {American Physical Society},
  doi = {10.1103/PhysRevB.84.205128},
  url = {https://link.aps.org/doi/10.1103/PhysRevB.84.205128}
}

@article{yao2018,
  title = {Edge States and Topological Invariants of Non-Hermitian Systems},
  author = {Yao, Shunyu and Wang, Zhong},
  journal = {Phys. Rev. Lett.},
  volume = {121},
  issue = {8},
  pages = {086803},
  numpages = {8},
  year = {2018},
  month = {Aug},
  publisher = {American Physical Society},
  doi = {10.1103/PhysRevLett.121.086803},
  url = {https://link.aps.org/doi/10.1103/PhysRevLett.121.086803}
}

@article{kadanoff1966,
  title = {Scaling laws for ising models near ${T}_{c}$},
  author = {Kadanoff, Leo P.},
  journal = {Physics Physique Fizika},
  volume = {2},
  issue = {6},
  pages = {263--272},
  numpages = {10},
  year = {1966},
  month = {Jun},
  publisher = {American Physical Society},
  doi = {10.1103/PhysicsPhysiqueFizika.2.263},
  url = {https://link.aps.org/doi/10.1103/PhysicsPhysiqueFizika.2.263}
}

@article{wilson1971,
  title = {Renormalization Group and Critical Phenomena. I. Renormalization Group and the Kadanoff Scaling Picture},
  author = {Wilson, Kenneth G.},
  journal = {Phys. Rev. B},
  volume = {4},
  issue = {9},
  pages = {3174--3183},
  numpages = {0},
  year = {1971},
  month = {Nov},
  publisher = {American Physical Society},
  doi = {10.1103/PhysRevB.4.3174},
  url = {https://link.aps.org/doi/10.1103/PhysRevB.4.3174}
}

@article{wilson1972,
  title = {Renormalization Group and Critical Phenomena. II. Phase-Space Cell Analysis of Critical Behavior},
  author = {Wilson, Kenneth G.},
  journal = {Phys. Rev. B},
  volume = {4},
  issue = {9},
  pages = {3184--3205},
  numpages = {0},
  year = {1971},
  month = {Nov},
  publisher = {American Physical Society},
  doi = {10.1103/PhysRevB.4.3184},
  url = {https://link.aps.org/doi/10.1103/PhysRevB.4.3184}
}

@book{wegner2016,
  title = {Supermathematics and Its {{Applications}} in {{Statistical Physics}}: {{Grassmann Variables}} and the {{Method}} of {{Supersymmetry}}},
  author = {Wegner, Franz},
  year = {2016},
  publisher = {Springer},
  isbn = {978-3-662-49170-6},
  langid = {english}
}

@article{kawabata2021,
  title = {Nonunitary Scaling Theory of Non-Hermitian Localization},
  author = {Kawabata, Kohei and Ryu, Shinsei},
  journal = {Phys. Rev. Lett.},
  volume = {126},
  issue = {16},
  pages = {166801},
  numpages = {7},
  year = {2021},
  month = {Apr},
  publisher = {American Physical Society},
  doi = {10.1103/PhysRevLett.126.166801},
  url = {https://link.aps.org/doi/10.1103/PhysRevLett.126.166801}
}

@article{hatsugai1997,
  title = {Disordered critical wave functions in random-bond models in two dimensions: Random-lattice fermions at $E=0$ without doubling},
  author = {Hatsugai, Yasuhiro and Wen, Xiao-Gang and Kohmoto, Mahito},
  journal = {Phys. Rev. B},
  volume = {56},
  issue = {3},
  pages = {1061--1064},
  numpages = {0},
  year = {1997},
  month = {Jul},
  publisher = {American Physical Society},
  doi = {10.1103/PhysRevB.56.1061},
  url = {https://link.aps.org/doi/10.1103/PhysRevB.56.1061}
}

@article{guruswamy2000,
  title = {gl($N$|$N$) Super-current algebras for disordered Dirac fermions in two dimensions},
  journal = {Nuclear Physics B},
  volume = {583},
  number = {3},
  pages = {475-512},
  year = {2000},
  issn = {0550-3213},
  doi = {https://doi.org/10.1016/S0550-3213(00)00245-5},
  url = {https://www.sciencedirect.com/science/article/pii/S0550321300002455},
  author = {S. Guruswamy and A. LeClair and A.W.W. Ludwig}
}

@article{horovitz2002,
  title = {Freezing transitions and the density of states of two-dimensional random Dirac Hamiltonians},
  author = {Horovitz, Baruch and Doussal, Pierre Le},
  journal = {Phys. Rev. B},
  volume = {65},
  issue = {12},
  pages = {125323},
  numpages = {10},
  year = {2002},
  month = {Mar},
  publisher = {American Physical Society},
  doi = {10.1103/PhysRevB.65.125323},
  url = {https://link.aps.org/doi/10.1103/PhysRevB.65.125323}
}

@article{mudry2003,
  title = {Density of states for the $\ensuremath{\pi}$-flux state with bipartite real random hopping only: A weak disorder approach},
  author = {Mudry, C. and Ryu, S. and Furusaki, A.},
  journal = {Phys. Rev. B},
  volume = {67},
  issue = {6},
  pages = {064202},
  numpages = {28},
  year = {2003},
  month = {Feb},
  publisher = {American Physical Society},
  doi = {10.1103/PhysRevB.67.064202},
  url = {https://link.aps.org/doi/10.1103/PhysRevB.67.064202}
}

@article{dellanna2006,
  title = {Anomalous dimensions of operators without derivatives in the non-linear {$\sigma$}-model for disordered bipartite lattices},
  journal = {Nuclear Physics B},
  volume = {750},
  number = {3},
  pages = {213-228},
  year = {2006},
  issn = {0550-3213},
  doi = {https://doi.org/10.1016/j.nuclphysb.2006.05.022},
  url = {https://www.sciencedirect.com/science/article/pii/S055032130600438X},
  author = {Luca Dell'Anna}
}

@article{markos2007,
  title = {Critical conductance of two-dimensional chiral systems with random magnetic flux},
  author = {Marko\ifmmode \check{s}\else \v{s}\fi{}, P. and Schweitzer, L.},
  journal = {Phys. Rev. B},
  volume = {76},
  issue = {11},
  pages = {115318},
  numpages = {8},
  year = {2007},
  month = {Sep},
  publisher = {American Physical Society},
  doi = {10.1103/PhysRevB.76.115318},
  url = {https://link.aps.org/doi/10.1103/PhysRevB.76.115318}
}

@article{motrunich2002,
  title = {Particle-hole symmetric localization in two dimensions},
  author = {Motrunich, Olexei and Damle, Kedar and Huse, David A.},
  journal = {Phys. Rev. B},
  volume = {65},
  issue = {6},
  pages = {064206},
  numpages = {17},
  year = {2002},
  month = {Jan},
  publisher = {American Physical Society},
  doi = {10.1103/PhysRevB.65.064206},
  url = {https://link.aps.org/doi/10.1103/PhysRevB.65.064206}
}

@article{bocquet2003,
  title = {Network models for localization problems belonging to the chiral symmetry classes},
  author = {Bocquet, Marc and Chalker, J. T.},
  journal = {Phys. Rev. B},
  volume = {67},
  issue = {5},
  pages = {054204},
  numpages = {13},
  year = {2003},
  month = {Feb},
  publisher = {American Physical Society},
  doi = {10.1103/PhysRevB.67.054204},
  url = {https://link.aps.org/doi/10.1103/PhysRevB.67.054204}
}

@article{Alexander2006KPM,
  title = {The kernel polynomial method},
  author = {Wei\ss{}e, Alexander and Wellein, Gerhard and Alvermann, Andreas and Fehske, Holger},
  journal = {Rev. Mod. Phys.},
  volume = {78},
  issue = {1},
  pages = {275--306},
  numpages = {0},
  year = {2006},
  month = {Mar},
  publisher = {American Physical Society},
  doi = {10.1103/RevModPhys.78.275},
  url = {https://link.aps.org/doi/10.1103/RevModPhys.78.275}
}

@article{lanczos1950iteration,
  title={An iteration method for the solution of the eigenvalue problem of linear differential and integral operators},
  author={Lanczos, Cornelius},
  journal={Journal of research of the National Bureau of Standards},
  volume={45},
  number={4},
  pages={255--282},
  year={1950}
}

@article{jackson1912approximation,
  title={On approximation by trigonometric sums and polynomials},
  author={Jackson, Dunham},
  journal={Transactions of the American Mathematical society},
  volume={13},
  number={4},
  pages={491--515},
  year={1912},
  publisher={JSTOR}
}

@Inbook{Skilling1989MaximumEntropy,
  author = {Skilling, John"},
  editor= {Skilling, J.},
  title={The Eigenvalues of Mega-dimensional Matrices"},
  bookTitle={Maximum Entropy and Bayesian Methods: Cambridge, England, 1988},
  year={1989},
  publisher={Springer Netherlands},
  address={Dordrecht},
  pages={455--466},
  isbn={978-94-015-7860-8},
  doi={10.1007/978-94-015-7860-8_48},
  url={https://doi.org/10.1007/978-94-015-7860-8_48}
}

@article{Drabold1993MaximumEntropy,
  title = {Maximum entropy approach for linear scaling in the electronic structure problem},
  author = {Drabold, David A. and Sankey, Otto F.},
  journal = {Phys. Rev. Lett.},
  volume = {70},
  issue = {23},
  pages = {3631--3634},
  numpages = {0},
  year = {1993},
  month = {Jun},
  publisher = {American Physical Society},
  doi = {10.1103/PhysRevLett.70.3631},
  url = {https://link.aps.org/doi/10.1103/PhysRevLett.70.3631}
}

@article{Silver1994Stochastic,
  author = {SILVER, R.N. and R\"{O}DER, H.},
  title = {DENSITIES OF STATES OF MEGA-DIMENSIONAL HAMILTONIAN MATRICES},
  journal = {International Journal of Modern Physics C},
  volume = {05},
  number = {04},
  pages = {735-753},
  year = {1994},
  doi = {10.1142/S0129183194000842}
}

@article{Nayak2024,
  title = {Band-center metal-insulator transition in bond-disordered graphene},
  author = {Nayak, Naba P. and Sarkar, Surajit and Damle, Kedar and Bera, Soumya},
  journal = {Phys. Rev. B},
  volume = {109},
  issue = {3},
  pages = {035109},
  numpages = {9},
  year = {2024},
  month = {Jan},
  publisher = {American Physical Society},
  doi = {10.1103/PhysRevB.109.035109},
  url = {https://link.aps.org/doi/10.1103/PhysRevB.109.035109}
}

\end{document}